\documentclass[prx,twocolumn,amsmath,amssymb,eqsecnum,nofootinbib,superscriptaddress
]{revtex4-2}
\usepackage{graphicx,amsmath,relsize,epstopdf,color,mathtools,bm,newtxtext,newtxmath,braket,rotating}
\usepackage[hyphenbreaks]{breakurl}
\usepackage[colorlinks=true,linkcolor=blue,citecolor=blue,urlcolor =blue]{hyperref}
\usepackage[normalem]{ulem}

\usepackage{comment}
\usepackage{booktabs}
\usepackage{float}

\usepackage[table,xcdraw]{xcolor}
\newcommand{\eq}[1]{\begin{equation}\begin{aligned}#1\end{aligned}\end{equation}}

\newcommand{\blue}[1]{\textcolor{blue}{#1	}}

\newcommand{\Cov}{\mathop{\mathrm{Cov}} \nolimits}
\newcommand{\Var}{\mathop{\mathrm{Var}} \nolimits}

\newcommand{\Tr}{\mathop{\mathrm{Tr}} \nolimits}

\newcommand{\eu}{\mathrm{e}}
\newcommand{\iu}{\mathrm{i}}

\newcommand{\sugg}[1]{#1}

\newcommand{\SCS}{\mathcal{A}}
\newcommand{\dd}{\mathrm{d}}

\usepackage{soul,xcolor}
\usepackage{dsfont}

\newenvironment{purpleblock}
{}
{}

\begin{document}
	
\setstcolor{red}

\title{Spin coherence scale: operator-ordering sensitivity beyond the Heisenberg-Weyl group}

\begin{abstract}
  We introduce the spin coherence scale as a measure of quantum coherence for spin systems, generalizing the quadrature coherence scale (QCS) previously defined for quadrature observables. 
  This SU($2$)-invariant measure quantifies the off-diagonal coherences of a quantum state in angular momentum bases, weighted by the classical distinguishability of the superposed states.
  It serves as a witness of nonclassicality\sugg{,}  provides both upper and lower bounds on the Hilbert–Schmidt distance to the set of classical (spin coherent) states\sugg{, and bounds the Wigner negativity of a spin state}. We demonstrate that many hallmark properties of the QCS carry over to the spin setting, including its links to noise susceptibility of a state and moments of quasiprobability distributions \sugg{and its experimental realizability with a two-copy scheme}. 
  The spin coherence scale has direct implications for quantum metrology in the guise of rotation sensing.
  We also generalize the framework to SU($n$) systems, identifying the unique SU($n$)-invariant depolarization channel and outlining a broad, Lie-algebraic approach to defining and characterizing the properties of coherence scale beyond harmonic oscillators.

\end{abstract}

\author{Aaron Z. Goldberg}
\affiliation{National Research Council of Canada, 100 Sussex Drive, Ottawa, Ontario K1N 5A2, Canada}
\author{Andre J. Rodrigues}
\affiliation{Department of Physics  and  Centre for Quantum Information and  Quantum Control$,$  University of Toronto$,$ 60 St George St$,$ Toronto$,$ Ontario$,$ M5S 1A7$,$ Canada}
\author{Y. Batuhan Yilmaz}
\affiliation{Department of Physics  and  Centre for Quantum Information and Quantum Control$,$  University of Toronto$,$ 60 St George St$,$ Toronto$,$ Ontario$,$ M5S 1A7$,$ Canada}
\author{Aephraim M. Steinberg}
\affiliation{Department of Physics and Centre for Quantum Information and Quantum Control$,$  University of Toronto$,$ 60 St George St$,$ Toronto$,$ Ontario$,$ M5S 1A7$,$ Canada} 
\author{Anaelle Hertz}
\affiliation{National Research Council of Canada, 100 Sussex Drive, Ottawa, Ontario K1N 5A2, Canada}

\maketitle
\tableofcontents

\section{Introduction}
Coherence and ambiguity have opposite meanings in English but may coincide in quantum physics. For to be coherent means a system has a propensity for interference~\cite{MandelWolf1995,BornWolf1999} between two distinguishable states that may be at odds with each other, as parodied by Schr\"odinger's cat~\cite{Schrodinger1935cat}. This is particularly true for quantum coherence, where the more diametrically opposed the terms in a superposition, the more quantum a state may be deemed~\cite{Aberg2006arxiv,Baumgratzetal2014}, leading to the ``coherence scale'' as a measure of coherence weighted by macroscopic distinguishability~\cite{Hertz}.

To this may be added another ingredient, that quantum states cannot provide simultaneous, exact values for the expectation values of noncommuting operators. In the case of position and momentum, this leads to the quadrature coherence scale (QCS) as a measure of the total coherence for any pair of maximally noncommuting quadrature observables; unsurprisingly, this quantity is related to noise properties of a state \cite{Hillery3}, a state's usefulness for multiparameter displacement sensing \cite{Goldbergetal2021rotationspublished}, loss properties of a state \cite{LupuGladsteinetal2025}, decoherence \cite{Hertz}, quasiprobability distribution fluctuations and quantum chaos \cite{GongBrumer}, and more. What is the appropriate coherence scale for operators with more complicated commutation relations
?

We begin with spin-$J$ systems, where we can define angular momentum operators in three orthogonal directions, denoted by $i=1$, $2$, and $3$, that satisfy the SU(2) 
commutation relations
\eq{
[J_1,J_2]=\iu J_3
\label{eq:SU2 commutation}
}
and cyclic permutations thereof ($\hbar=1$). This is the regime of optical polarization~\cite{Goldbergetal2021polarization}, structured light~\cite{DennisAlonso2017}, Bose-Einstein condensates~\cite{Albiezetal2005}, and any other physical system whose mathematics resembles angular momentum, such as magnets. A quantum state \sugg{$\rho$} will commute with an angular momentum operator if the former is an eigenstate of the latter, while it will maximally noncommute when it is a maximally coherent superposition of two angular momentum eigenstates. With this in mind, a simple definition of the spin coherence scale is the readily calculable
\eq{
\SCS^2\propto \sum_{i=1}^3\Tr([\rho,J_i]^2).
} 
We then proceed to study $\SCS^2$ and its properties.

First, we show that $\SCS^2$ has appropriate behaviour for maximally and minimally noncommutative states. We link its value to the usefulness of a state for sensing rotations in three dimensions, which has garnered recent attention in quantum metrology \cite{KolenderskiDemkowiczDobrzanski2008,Stocktonetal2011,ChryssomalakosHC2017,Bouchardetal2017,Martinetal2020}. Just as for the QCS, $\SCS^2$ is exactly a measure of the coherence present in a density operator expressed in an angular momentum basis, weighted by how classically different the two states being superposed are, then averaged over all three angular momentum operators. \sugg{The angular momentum bases are comprised of eigenstates of the angular momentum operators, $J_i|J,m\rangle_i=m|J,m\rangle_i$, where $m\in\{-J,-J+1,\cdots,J\}$. The spin coherence scale then takes the form}
\eq{
\SCS^2\propto\sum_{i=1}^3\sum_{m,m^\prime}(m-m^\prime)^2|\vphantom{a}_i\langle \sugg{J,}m|\rho|\sugg{J,}m^\prime\rangle_i|^2.
} 
It is SU(2) invariant so that the orientation of the spin is immaterial, as desired, and linked to numerous measures of quantumness for spins \cite{Goldbergetal2020extremal}. Moreover, $\SCS^2$ provides an upper and lower bound on distance measures to the set of classical states\sugg{; i.e., to all states that can be written as a convex mixture of coherent states \cite{Titulaer,Richter}.}
It is a \textit{bona fide} witness of nonclassicality,\footnote{\sugg{The meaning of ``classicality'' is not unanimous. Here, we term ``classical'' any state written as a convex mixture of coherent states; viz., states that have positive Glauber-Sudarshan-type quasiprobability distributions \cite{Titulaer}. Note that witnesses of nonclassicality for continuous-variable states other than the QCS are also possible, with famous criteria including  the Wigner negativity \cite{Kenfack}, Wigner-Yanasee skew information \cite{Luo}, Mandel Q parameter and degree of squeezing \cite{Agarwal2,Richter,Ryl}, the entanglement potential \cite{Asboth}, and the Kirkwood-Dirac distribution \cite{DeBievre23,ArvidssonShukur2024KD}. When the same witnesses are applied to spin systems, they do not always agree as to which states are the most quantum~\cite{Goldbergetal2020extremal}.}} and exhibits all of the intriguing connections to decoherence and noise properties of a state that are found for the QCS~\cite{CompanionLongArxiv}, where now depolarization noise is more appropriate than loss for this physical system. The connection to quasiprobability distribution fluctuations is more challenging due to quasiprobabilities having richer properties for SU(2) than for the more familiar Heisenberg-Weyl group, but we show that many of the QCS-style properties hold and many more hold in the limit of large spins.

\sugg{The spin coherence scale is similar to other measures of quantumness for spin systems in that it quantifies how different a state is from a spin-coherent state,
where such measures have been introduced using anticoherence~\cite{Zimba2006,Baguetteetal2015,Giraudetal2015,BaguetteMartin2017} and the Majorana representation~\cite{Hannay1998JPA,UshaDevietal2012,Bruno2012,Yangetal2015,Bjorketal2015PRA,LiuFu2016,ChryssomalakosHC2017,Chryssomalakosetal2018}, the Wehrl entropy~\cite{Lee1988,LiebSolovej2014,BaecklundBengtsson2014}, distance measures~\cite{Giraudetal2010}, entanglement properties~\cite{Stocktonetal2003,Hubeneretal2009,TothGuhne2009,Kieseletal2010,Martinetal2010,Aulbachetal2010,RibeiroMosseri2011,Markham2011,Baguetteetal2014,Sperlingetal2019,Goldbergetal2022sym}, spin squeezing~\cite{KitagawaUeda1993,LuisKorolkova2006,Klimovetal2010,Maetal2011,Jingetal2019}, and more~\cite{delaHozetal2013,Goldbergetal2020extremal}.
A conspicuously absent quantity from this list is the notion of Wigner negativity, which is a common measure of nonclassicality for continuous-variable systems~\cite{Kenfack} and provides an essential ingredient for quantum computation~\cite{Veitchetal2012} but is notoriously difficult to employ for spin systems~\cite{Davisetal2021}. This is important because Wigner negativity is equivalent to contextuality for spin systems~\cite{Delfosseetal2017} and there contextuality provides the magic for quantum computation~\cite{Howardetal2014}. Fortunately, our spin coherence scale can be used to bound the Wigner negativity in a spin system, directly providing a method for assessing classicality.}

\sugg{A final important aspect of any measure or scale is its experimental viability. We provide a scheme for experimentally determining the spin coherence scale for optical systems using readily available laboratory components, which certifies its readiness for application.}

Given these bountiful parallels that arise when we define $\SCS^2$ analogously to the QCS, it is natural to ask whether the same extension can be done for any Lie group or any set of noncommuting operators. We briefly outline how these results can apply to any of the important groups SU($n$), with the only caveats being the special properties of quasiprobability distributions that remain open questions in the mathematical physics literature.

\section{Background: QCS properties}
To set the stage for numerous parallels, we recapitulate the definition and properties of the QCS \cite{Debievre,Hertz}. Consider two quadrature operators such as position and momentum satisfying the canonical commutation relations of the Heisenberg-Weyl group
\eq{[x,p]=\iu,
\label{eq:xp commutation}
} alternatively expressed using bosonic creation and annihilation operators $x=(a+a^\dagger)/\sqrt{2}$ and $p=-\iu(a-a^\dagger)/\sqrt{2}$ obeying $[a,a^\dagger]=1$. The QCS has equivalent definitions
\eq{
\mathcal{C}^2(\rho)
&=\frac{1}{2}\int dx dx^\prime (x-x^\prime)^2P_\rho(x,x^\prime)\\
&\,+\frac{1}{2}\int dp dp^\prime (p-p^\prime)^2P_\rho(p,p^\prime)\\
&=\frac{1}{2\mathcal{P}(\rho)}\Tr([\rho,x][x,\rho]+[\rho,p][p,\rho])\\
&=\frac{1}{\mathcal{P(\rho)}}\Tr(\rho(-2a\rho a^\dagger+a^\dagger a\rho+\rho a^\dagger a))+1
\label{eq:QCS defns}
} where $P_\rho(x,x^\prime)= |\langle x|\rho|x^\prime\rangle|^2/\mathcal{P}(\rho)$ is a unit-normalized probability distribution, $\mathcal{P}(\rho)=\Tr(\rho^2)$ is the purity of a state $\rho$,  $|x\rangle$ is an eigenstate of $x$, and analogous definitions hold for momentum.\footnote{\sugg{This may be contrasted with the Wigner-Yanase skew information of a state with respect to quadrature operators, where $\rho/\mathcal{P}(\rho)$ is replaced by $\sqrt{\rho}$~\cite{LuoZhang2019} so that it does not directly measure the coherences $|\langle x|\rho|x^\prime\rangle|$ but the related quantity $|\langle x|\sqrt{\rho}|x^\prime\rangle|$ that coincides for pure states.}} The larger the QCS, the more coherence a state has, with $\mathcal{C}^2(\rho)>1$ certifying nonclassicality and all states obeying $\mathcal{C}^2(\rho) > 0$.

A standard definition of coherence is the magnitude of the off-diagonal elements of a density matrix in some basis; this is the function of $P_\rho(x,x^\prime)$, which turns these magnitudes into a probability distribution. Since coherence is a basis-dependent quantity, there is no physical grounds on which to single out position or momentum, so we average over the two. And, since even the orientation of the quadratures is rather immaterial, the QCS boasts the property of being invariant to phase-space rotations, changing not at all when the quadratures are rotated as $x\to x\cos\theta +p\sin\theta $ and $p\to -x\sin\theta +p\cos\theta $. Finally, the magnitude of the coherence $|\langle x|\rho|x^\prime\rangle|$ is weighted by the distance between the eigenvalues $|x-x^\prime|$ because, even within a chosen basis, the order of the eigenstates has physical meaning. This provides a scale for the coherence in the quadratures, hence the name QCS. 

Next, this coherence scale can be expressed as the degree to which a state commutes with the quadrature operators. To find the size of the noncommutativity of terms like $[x,\rho]$, we cannot simply take the trace because it vanishes, hence the appearance of $-[\rho,x]^2$. Because the QCS applies equal weight to position and momentum, a position eigenstate $\rho=|x\rangle\langle x|$ that commutes with $x$ will still have large QCS, given that it does not commute with $p$\sugg{; this is approximately true for any physical state that approximates a position eigenstate, where a pure state that mostly commutes with one quadrature operator will not commute with an orthogonal quadrature operator}. The normalization by $\mathcal{P}(\rho)$ then ensures proper behaviour since some commutativity may come from the mixedness of a state. Then, a rewriting of the QCS in terms of creation and annihilation operators yields the expectation value with respect to $\rho$ of the rate of change of a state in a loss or an amplitude damping channel $\partial \rho/\partial t=a\rho a^\dagger-(a^\dagger a\rho+\rho a^\dagger a)/2$, which proves useful \cite{Hertz}.

These second and third expressions allow for many manipulations. For a pure state \sugg{$\ket{\psi}$},
\eq{
\mathcal{C}^2(\sugg{\ket{\psi}})=\Var_\psi(x)+\Var_\psi(p)=\langle a^\dagger a\rangle_\psi-|\langle a\rangle_\psi|^2+1,
\label{eq:QCS pure}
} where $\Var_\rho(X)=\langle X^2\rangle_\rho-\langle X\rangle_\rho^2$. The middle expression of Eq.~\eqref{eq:QCS pure} is sometimes called the total noise of the state and is equivalent to the direction-averaged quantum Fisher information for a state that is used for sensing displacements in phase space~\cite{Goldbergetal2020extremal} \sugg{and to the Wigner-Yanase skew information of a state with respect to $a$ or the average of such informations with respect to two orthogonal quadratures~\cite{LuoZhang2019} (which may also be seen trivially from Eq.~\eqref{eq:QCS defns} with the pure-state condition $\rho=\sqrt{\rho}$)}. The right-hand side, in turn, shows that only coherent states with $a|\alpha\rangle\propto |\alpha\rangle$ exist on the classical/quantum threshold of $\mathcal{C}^2(\rho)=1$, while all other pure states are decidedly nonclassical with $\mathcal{C}^2(\rho)>1$. Then, all classical states, defined as convex combinations of coherent states, obey $\mathcal{C}^2(\rho)\leq 1$, which is why $\mathcal{C}^2(\rho)>1$ certifies quantumness. This is cemented in the relationship between the QCS and the distance $\mathcal{D}(\rho)$ to the set of classical states according to the Hilbert-Schmidt norm \sugg{\cite{Debievre}}, where the former bounds the latter from both sides:
\eq{
\mathcal{C}(\rho)-1\leq \mathcal{D}(\rho)\leq \mathcal{C}(\rho).
}

The QCS can be measured with two copies of a state, a beam splitter, and a photon-number-resolving detector~\cite{Griffet,Goldbergetal2023}. Because this setup is intimately connected to noise in linear optics~\cite{CavesCrouch1987,Jeffersetal1993,NielsenChuang2000,Leonhardt2003}, where interference with a later-ignored vacuum at a beam splitter is the dominant loss model and interference with a thermal state a more sophisticated origin of decoherence, it is reasonable that the QCS is related to these dynamical processes via \cite{CompanionLongArxiv, Hertz}
\eq{
\mathcal{C}^2(\rho_T)=\frac{T}{\mathcal{P}(\rho_T)}\frac{\partial \mathcal{P}(\rho_T)}{\partial T}+1,
} where $T$ is the transmittance of the beam splitter and $\rho_T$ the state after loss $1-T$.\footnote{\sugg{Equivalently,  $\rho_T=\Tr_b[B(T)(\rho_1\otimes\ket{0}\bra{0})B^\dag(T)]$ where $B(T)=\text{e}^{(a b^\dagger-b^\dagger a)\arccos\sqrt{T}}$ is the beam splitter unitary that mixes two bosonic modes annihilated by $a$ and $b$, $\rho_1$ is any initial state, and the partial trace over the second mode leads ensures that those photons are lost. Also equivalently, $\rho_T$ is any quantum state evolving under the aforementioned amplitude-damping channel with $T=e^{-t}$~\cite{WeidlichHaake1965,KarrleinGrabert1997,ScullyZubairy1997,GardinerZoller2004}.}}
The QCS is thus a measure of a system's rate of change of purity or of its susceptibility to noise; the more fragile a state, the more quantum it is. \sugg{Any state whose purity decreases with loss will be witnessed to be nonclassical, in contrast with coherent states whose purities are unchanged by loss and with classical combinations of coherent states whose purities increase with loss.} It was recently shown that this purity is convex in beam splitter transmission and thus that the QCS decays monotonically from certifying to not certifying quantumness when a pure state undergoes loss, with evidence and conjectures that this is true for all states~\cite{CompanionLongArxiv}. \sugg{Because purity after a beam splitter is an entanglement monotone, the QCS connects to the ``entanglement potential''~\cite{Asbothetal2005} of a state and was used to prove an old conjecture about how to best generate entanglement with linear optics~\cite{LupuGladsteinetal2025}.}

Finally, the QCS enjoys many relations to quasiprobability distributions due to the above properties. For example, the $s$-ordered quasiprobability distributions \cite{CahillGlauber}
\eq{
W^{(s)}_\rho(\alpha)=\frac{1}{\pi^2}\int \dd^2\beta \ e^{s\frac{|\beta |^2}{2}+\beta ^*\alpha -\beta \alpha ^*}\Tr[\rho D(\beta )],
} with the displacement operator $D(\beta)=\exp(\beta a^\dagger-\beta^* a)$ acting on coherent states as $D(\beta )|\alpha \rangle=|\alpha +\beta \rangle$,
all lead to expressions for a state's purity via
\eq{
\mathcal{P}(\rho)={\pi}\int d^2\alpha W^{(-s)}_\rho(\alpha)W^{(s)}_\rho(\alpha),
\label{eq:purity from quasi CV}
} where $s=1$, $0$, and $-1$ correspond to the famous Glauber-Sudarshan, Wigner, and Husimi functions, respectively. Due to the correspondence principles for how states evolving as $\rho\to a\rho$ affect the quasiprobability distributions, or, equivalently, due to how quasiprobability distributions relate to each other when states lose photons, the QCS can be expressed in many forms; e.g. \cite{Hertz,CompanionLongArxiv},
\begin{eqnarray}\label{eq:QCS quasi}
\mathcal{C}^2(\rho)&=&{1-}\frac{1}{\mathcal{P}(\rho)}\! \int \!d^2\alpha d^2\beta W_\rho^{(1)}(\alpha) W_\rho^{(1)}(\beta)\eu^{-|\alpha-\beta|^2}|\alpha-\beta|^2\nonumber\\
&=&\frac{\int d^2\alpha\boldsymbol{\nabla} W_\rho^{(0)}(\alpha)\cdot\boldsymbol{\nabla} W_\rho^{(0)}(\alpha)  }{ {4}\int  d^2\beta\left( W_\rho^{(0)}(\beta)\right)^2}.
\end{eqnarray}
The final expression for the divergence of the Wigner function found earlier use in studies of quantum chaos~\cite{Gu1990,GongBrumer}, showcasing the diverse realm of influence of the QCS.

\section{Definitions of spin coherence scale coincide}
We begin our exegesis with a rigorous definition of the spin coherence scale to showcase how all analogous definitions {\`a la QCS} hold true. 

First, we consider the eigenbases of the three angular momentum generators, writing them as $J_i|J,m\rangle_i=m|J,m\rangle_i$; the generators are Hermitian. In practice, we should specify a total spin $J$ that derives from the Casimir operator $\mathbf{J}^2=\sum_{i=1}^3 J_i^2=J(J+1)$, which may be considered a positive half-integer for a fixed total spin or an operator otherwise. Then $m\in\{-J,-J-1,\cdots,J\}$, the phase space is a sphere, and rotations of the sphere are generated by the unitary operators
\eq{
R(\theta,\mathbf{n})=\exp(\iu \theta\mathbf{J}\cdot\mathbf{n});\quad \mathbf{J}=(J_1,J_2,J_3)^\top
\label{eq:rot op}
} that rotate the angular momentum operators by $\theta$ about axis $\mathbf{n}$ when applied as $RJ_i R^\dagger$. The action on states, for example, lets us identify the eigenstates of each operator as rotated versions of eigenstates of the other operators via relations like $|J,m\rangle_1=R(\pi/2,(0,1,0)^\top )|J,m\rangle_3$. With this machinery, the spin coherence scale can be defined as the manifestly positive
\eq{
\SCS^2(\rho)=\frac{1}{{2}J}\sum_{i=1}^3 \sum_{m,m^\prime=-J}^J (m-m^\prime)^2 P_\rho(m,m^\prime;i).
\label{eq:coherence prob dist}
} Here, since the purity can be written as $\mathcal{P}(\rho)=\sum_{m,m^\prime=-J}^J|\vphantom{a}_i\langle J,m|\rho|J,m^\prime\rangle_i|^2$ for all $i$, the unit-normalized probability distribution \eq{
P_\rho(m,m^\prime;i)=\frac{|\vphantom{a}_i\langle J,m|\rho|J,m^\prime\rangle_i|^2}{\mathcal{P}(\rho)}
} encodes the magnitudes of all of the off-diagonal elements of the density matrix in the eigenbasis of the $i$th angular momentum operator and the sum is discrete. Again, to make this a coherence scale, the coherence is weighted by the magnitude of $m-m^\prime$ and averaged over all orientations. We will later show it to be unchanged by SU(2) transformations of the state; i.e., rotations. The normalization by $J$ is convenient because the SU(2) noncommutativity [Eq.~\eqref{eq:SU2 commutation}] grows with $J$ in contrast to the Heisenberg-Weyl (position-momentum) noncommutativity [Eq.~\eqref{eq:xp commutation}].

This definition is exactly the same as
\eq{
\SCS^2(\rho)=\frac{1}{{2}J\mathcal{P}(\rho)}\sum_{i=1}^3\Tr([\rho,J_i][J_i,\rho]).
\label{eq:SCS commutators}
} To see this, resolve the identity between the commutators and evaluate the trace in the eigenbasis of the $i$th angular momentum operator for the $i$th term in the sum.
{For pure states, the spin coherence scale reduces to 
\eq{
\SCS^2(|\psi\rangle)=\frac{1}{J}\sum_{i=1}^3\Var_\psi(J_i).
\label{SCSpure}}}

The expression in Eq.~\eqref{eq:SCS commutators} readily leads to SU(2) invariance by noting {that $\Tr([\rho,J_i][J_i,\rho])=2\Tr(J_i^2\rho^2-(J_i\rho)^2)$ and} the independent invariances of $\Tr(\rho^2)$, $\sum_{i=1}^3 J_i^2$, and $\sum_{i=1}^3\Tr((J_i\rho)^2)$ due to \eq{
RJ_iR^\dagger=\sum_{j=1}^3 r_{ij}J_j
\label{eq:rotation of Ji}
} for an orthogonal matrix with elements $r_{ij}$, but another expression discharges our burden of proof. A tiny rearrangement leads to 
\eq{
\SCS^2(\rho)=-\frac{1}{2\mathcal{P}(\rho)}\frac{\partial \mathcal{P}(\rho)}{\partial t}=-\frac{1}{2}\frac{\partial \ln \mathcal{P}(\rho)}{\partial t}
\label{eq:SCSasderivativePurity}} for a state evolving under the isotropic depolarization channel \cite{RivasLuis2013}
\eq{
\frac{\partial \rho}{\partial t}=\frac{1}{J}\sum_{i=1}^3 {\left(J_i \rho J_i-\frac{J_i^2\rho+\rho J_i^2}{2}\right)=-\frac{1}{2J}\sum_{i=1}^3[J_i,[J_i,\rho]]}.
\label{eq:depol Lindblad}
} This is a unital, Lindbladian evolution with Lindblad operators $J_i$ and was demonstrated to be invariant under rotations in Ref.~\cite{RivasLuis2013}, where it was given a physical origin in terms of randomness and models the dominant noise source for optical polarization, relevant also for qubits and magnetic sublevels. In contrast to the QCS and the Heisenberg-Weyl group, where the dominant source of noise is photon loss and the sole steady state is the vacuum (a pure, coherent state), for spins the sole steady state is the maximally mixed one. Once depolarization is chosen as the noise model, $\mathcal{C}^2$ and $\SCS^2$ become highly analogous and provide the immediate identification of
\eq{
\SCS^2(R(\theta,\mathbf{n})\rho R(\theta,\mathbf{n})^\dagger)=\SCS^2(\rho).
}

\sugg{Finally, it is instructive to consider the large-spin limit. Using the Holstein--Primakoff transformation \cite{HolsteinPrimakoff1940}, the spin operators may be mapped to a single bosonic mode $a$, and when the excitation number satisfies $a^\dagger a \ll J$, the SU(2) algebra approaches the Heisenberg--Weyl algebra. This condition corresponds to states that are localized near a classical spin direction on the Bloch sphere. In this regime, the spin coherence scale reduces to the quadrature coherence scale up to $\mathcal{O}(1/J)$ corrections. This establishes consistency with continuous-variable notions of nonclassicality and shows that the SCS provides a natural finite-dimensional generalization of the QCS.}

\sugg{
To gain an intuition for the spin coherence scale, consider the so-called spin-cat states in the $J_3$ basis~\cite{Sanders1989}
\eq{|\mathrm{cat}\rangle=\frac{\ket{J,n}_3+\ket{J,-n}_3}{\sqrt{2}},\quad n\geq 0,}
which are related to NOON states and GHZ states when $n=J$~\cite{SandersGerry2014}. For these states and this basis, there is a single coherence $P_{\mathrm{cat}}(n,-n;3)=1/4$ that scales as $4n^2$, agreeing with the challenge of constructing spin-cat states with larger $n$~\cite{Yuetal2025}. Larger $n$ yields greater quantum advantages in metrology~\cite{Goldbergetal2021rotationspublished}. Further, when the spin-$J$ systems are comprised of symmetric states of $2J$ qubits, larger $n$ implies a superposition of more dissimilar qubits, with $J-n$ of the qubits being spin down, $J-n$ being spin up, and a superposition of the other $2n$ being all spin up or all spin down~\cite{Jonesetal2009,Simmonsetal2010}. According to the Majorana stellar representation, increasing $n$ decreases the degeneracy of the constellation and spreads the stars further about the sphere~\cite{Devietal2012}. In these senses the scale of the coherences is paramount and the spin coherence scale then averages the coherences over all angular momentum bases.}

\sugg{Mixedness also plays a role for this scale. Considering qubit states that are spins with $J=1/2$, which without loss of generality may be parametrized by their purity (because our scale is rotationally invariant), the spin coherence scale takes the value $\SCS^2(\rho)=2-1/\mathcal{P}(\rho)$, which decreases monotonically as the purity decreases. That maximally mixed states commute with all spin operators indicates that they achieve the minimal value $\SCS^2(\tfrac{\mathds{1}}{2J+1})=0$ in any dimension; when the spins are formed from light's polarization degree of freedom these are known as completely unpolarized light~\cite{Agarwal1971,PrakashChandra1971}.}

\section{Spin coherence scale as witness of quantumness}
In the domain of spins, the most classical states are the spin-coherent states\footnote{Since spin-coherent states are often abbreviated as ``SCS,'' we avoid this acronym throughout to alleviate conflict with ``spin coherence scale.'' These are also known as SU(2)-coherent states, atomic coherent states, and Bloch-coherent states.}~\cite{Radcliffe1971,Arecchietal1972,GlauberHaake1976,Agarwal1981,Perelomov1986,Giraudetal2008,Gazeau2009}. They are maximal eigenstates 
of angular momentum operators, satisfying $\mathbf{J}\cdot\mathbf{n}|\Omega^{(J)}\rangle=J|\Omega^{(J)}\rangle$, where $\Omega$ symbolizes the angular coordinates of the unit vector $\mathbf{n}$, and also satisfy $\mathbf{J}^2\ket{\Omega^{(J)}}=J(J+1)\ket{\Omega^{(J)}}$ due to having fixed total spin. The astute reader will realize that all such spin-coherent states are of the form $R(\theta,\mathbf{n})|J,J\rangle_i$ and will be happy to know they form an overcomplete basis for a fixed spin; this parallels the displacement and basis properties of canonical coherent states. The states are known as the most classical for a number of reasons~\cite{Goldbergetal2020extremal}, largely because they saturate various uncertainty relations such as~\cite{Delbourgo1977,HofmannTakeuchi2003,Dammeieretal2015,Dodonov2018}
\eq{
\sum_{i=1}^3\langle J_i\rangle_\rho^2 \leq \langle \mathbf{J}^2-J\rangle_\rho \quad\Rightarrow\quad \sum_{i=1}^3 \Delta^2 J_i\geq J
} 
and, for any three angular momentum operators satisfying the usual commutation relations, \eq{\Delta J_1\Delta J_2\geq\frac12|\langle J_3\rangle|.} 
In the context of optical polarization, this is equivalent to light's degree of polarization being upper bounded by unity~\cite{Luis2002}. Convex combinations of spin-coherent states \sugg{$\rho_{\mathrm{cl}}~=~\sum_k q_k |\Omega^{(J)}_k\rangle\langle \Omega^{(J)}_k|$ ($\sum_k q_k=1$, $q_k\geq 0$)} are again deemed classical \cite{Giraudetal2008}; flipping a coin to choose between classical states will not lead to quantumness. Stemming from the principle of quantumness being a measure of distinction from spin-coherent states, nonclassicality measures for spin systems have been introduced using anticoherence~\cite{Zimba2006,Baguetteetal2015,Giraudetal2015,BaguetteMartin2017} and the Majorana representation~\cite{Hannay1998JPA,UshaDevietal2012,Bruno2012,Yangetal2015,Bjorketal2015PRA,LiuFu2016,ChryssomalakosHC2017,Chryssomalakosetal2018}, the Wehrl entropy~\cite{Lee1988,LiebSolovej2014,BaecklundBengtsson2014}, distance measures~\cite{Giraudetal2010}, entanglement properties~\cite{Stocktonetal2003,Hubeneretal2009,TothGuhne2009,Kieseletal2010,Martinetal2010,Aulbachetal2010,RibeiroMosseri2011,Markham2011,Baguetteetal2014,Sperlingetal2019,Goldbergetal2022sym}, spin squeezing~\cite{KitagawaUeda1993,LuisKorolkova2006,Klimovetal2010,Maetal2011,Jingetal2019}, and more~\cite{delaHozetal2013,Goldbergetal2020extremal}\sugg{, as mentioned in the introduction}.

\sugg{The spin coherence scale provides the witness}
\eq{
\mathcal{A}^2(\rho)> 1\quad \Rightarrow\quad \rho\sugg{\neq \rho_{\mathrm{cl}}}\,\mathrm{\sout{quantum}};
} alternatively, for all classical states,  $\mathcal{A}^2(\sugg{\rho_{\mathrm{cl}}}){\leq} 1$.
By the above-quoted properties of spin-coherent states, we take the pure-state version (c.f. Eq.~\eqref{SCSpure})
to find $\SCS^2(|\Omega^{(J)}\rangle)=1$ and $\SCS^2(|\psi\rangle)>1$ for all $\psi\neq \Omega^{(J)}$; just like the QCS version with $\psi\neq \alpha$, all pure states other than coherent states are certifiably quantum. As for mixed states, we cannot immediately use convexity properties, because of the $\mathcal{P}(\rho)$ factor in the denominator. We also cannot use the quasiprobability diffusion technique used for proving the classical bound for the QCS in Ref. \cite{Debievre}. Instead, noting the cancellation in $\mathcal{P}(\rho)$, we identify
\eq{
\SCS^2(\rho)=J+1-\frac{\sum_{i=1}^3\Tr(J_i\rho J_i \rho)}{J\mathcal{P}(\rho)}.
\label{eq:SCS spin simple}
} Writing the most general classical state as $\rho_{\mathrm{cl}}~=~\sum_k q_k |\Omega^{(J)}_k\rangle\langle \Omega^{(J)}_k|$ for $\sum_k q_k=1$, $q_k\geq 0$, we are almost ready to prove $\SCS^2(\rho_{\mathrm{cl}})\leq 1$ for all classical states. The proof relies on a term-by-term inequality for all pairs of coherent states, which we establish by noting that $\sum_{i=1}^3 |\langle\Omega_k^{(J)}|J_i|\Omega_{k^\prime}^{(J)}\rangle|^2$ is invariant under rotations $J_i\to RJ_iR^\dagger$ (due to the orthogonality property in Eq.~\eqref{eq:rotation of Ji}), selecting the rotation that achieves $R^\dagger|\Omega_{k^\prime}^{(J)}\rangle=|J,J\rangle$ to elicit a maximal eigenvalue of $J_3$, then computing
\eq{
\sum_{i=1}^3 |\langle\Omega_k^{(J)}|J_i|\Omega_{k^\prime}^{(J)}\rangle|^2= \sum_{i=1}^2|\langle\Omega_k^{(J)}|RJ_iR^\dagger|\Omega_{k^\prime}^{(J)}\rangle|^2\\
+J^2|\langle\Omega_k^{(J)}|R|J,J\rangle|^2\geq J^2|\langle\Omega_k^{(J)}|\Omega_{k^\prime}^{(J)}\rangle|^2.
\label{eq:off diag coherent state inequality}
}
We then use Eq.~\eqref{eq:SCS spin simple} to prove the desired inequality for the spin coherence scale of classical states:
\eq{
\SCS^2(\rho_{\mathrm{cl}})=&J+1-\frac{\sum_{i=1}^3 \sum_{kk^\prime}q_k q_{k^\prime}|\langle \Omega^{(J)}_k| J_i|\Omega^{(J)}_{k^\prime}\rangle|^2}{J\sum_{kk^\prime}q_k q_{k^\prime}|\langle \Omega^{(J)}_k|\Omega^{(J)}_{k^\prime}\rangle|^2}\\
&\leq J+1-J\frac{ \sum_{kk^\prime}q_k q_{k^\prime}|\langle \Omega^{(J)}_k|\Omega^{(J)}_{k^\prime}\rangle|^2}{\sum_{kk^\prime}q_k q_{k^\prime}|\langle \Omega^{(J)}_k|\Omega^{(J)}_{k^\prime}\rangle|^2}= 1.
\label{eq:convex coherent inequality}
} For this we computed the purity to be $\mathcal{P}(\rho_\mathrm{cl})=\sum_{kk^\prime}q_k q_{k^\prime}|\langle \Omega^{(J)}_k|\Omega^{(J)}_{k^\prime}\rangle|^2$ and could only apply the inequality to the entire sum because each term had a negative coefficient $-q_k q_{k^\prime}$, which explicitly makes connection to the classical nature of the states' coefficients $q_k$.


When considering convex combinations of coherent states, we note that all single-qubit states are classical. All two-dimensional pure states are spin-coherent states and, therefore, any mixed state comprised thereof is encompassed by $\rho_{\mathrm{cl}}$\sugg{, as pointed out by Ref.~\cite{Giraudetal2008}}. This is reminiscent of the Kochen-Specker theorem only holding in dimensions greater than two such that, from the perspective of contextuality, all single-qubit states are again classical~\cite{Bell1966,KochenSpecker1967,Mermin1993}.

Our nonclassicality witness can then be used to bound the distance between a given state and the set of classical states. Following the procedure laid out for the QCS in Ref. \cite{Debievre}, we define an inner product by
\eq{
(A,B)=\frac{1}{{2}J}\sum_{i=1}^3\Tr([A^\dagger,J_i][J_i,B])
}
and thence the norm $|||A|||=\sqrt{(A,A)}$. With the Hilbert-Schmidt-normalized state $\tilde{\rho}=\rho/{\sqrt{\Tr(\rho^2)}}$, we can thus identify the spin coherence scale with the squared norm
\eq{
\SCS^2(\rho)=|||\tilde{\rho}|||^2.
} When $\rho$ is classical, $|||\tilde{\rho}|||^2\leq 1$; whereas, for nonclassical $\rho$, the measure
\eq{
\mathcal{D}(\rho)=\inf_{\rho_{\mathrm{cl}}}|||\tilde{\rho}-\tilde{\rho}_\mathrm{cl}|||
} encapsulates the distance between the state and the closest classical state and will be positive (other distance measures can also be considered~\cite{Giraudetal2008,Giraudetal2010}).
Since the set of states with $|||\tilde{\rho}|||^2\leq 1$ is convex (a unit ball) and has $\mathcal{D}(\rho)=0$, the triangle inequality for norms dictates the same identity as for the QCS:
\eq{
\SCS(\rho)-1\leq \mathcal{D}(\rho)\leq \SCS(\rho).
} 

{We prove the lower bound by writing 
\eq{|||\tilde{\rho}|||\leq|||\tilde{\rho}-\tilde{\rho}^*_{cl}|||+|||\tilde{\rho}^*_{cl}|||\leq\mathcal{D}(\rho)+1\nonumber} where $\rho^*_{cl}$ is the classical state the minimize the distance to the set of classical states.
For the upper bound, let us choose the maximally mixed state $\mathds{1}/(2J+1) $ for which $|||\mathds{1}/\sqrt{2J+1} |||=0$. Then 
$D\leq |||\tilde{\rho}-\mathds{1}/\sqrt{2J+1} |||\leq|||\tilde{\rho}|||+|||  \mathds{1}/\sqrt{2J+1} |||=|||\tilde{\rho}|||$.}
The spin coherence scale thus bounds the distance to the set of classical states and, when $\SCS^2\gg 1$, it almost exactly equals that distance.

It is natural to ask which states are the most quantum according to the spin coherence scale. For pure states, the answer is any state with $\langle J_i\rangle=0$ for all $i$:
\eq{
\SCS^2(|\psi_\mathrm{max}\rangle)=J+1.
}
These states, sometimes called first-order-unpolarized or $1$-anticoherent~\cite{Zimba2006}, have useful properties for metrology and can sometimes display ``hidden polarization'' in their higher-order moments~\cite{Klyshko1997,Bushevetal2001,Usachevetal2001}. \sugg{An example is the ``tetrahedron state'' $(\ket{2,2}+\sqrt{2}\ket{2,-1})/\sqrt{3}$~\cite{GoldbergJames2018Euler} or other ``Kings of quantumness''~\cite{Bjorketal2015,Bjorketal2015PRA} that have been experimentally generated in light's orbital angular momentum~\cite{Bouchardetal2017,Erikssonetal2023} and polarization~\cite{Ferrettietal2024} degrees of freedom.} Similar to the QCS, where pure states have their quantumness scale linearly with energy, the spin coherent scale grows linearly with the spin, which is the analogous property to the energy when considering the Casimir invariant or spins made from symmetric states of qubits. As for mixed states, the expression in Eq.~\eqref{eq:SCS spin simple} subtracts off the positive quantities $\Tr(J_i \rho J_i\rho)$, which are the sums of the squared singular values of the Hermitian operators $\sqrt{\rho}J_i\sqrt{\rho}$, again proving for all mixed states that
\eq{
\SCS^2\leq J+1.
} To saturate the inequality for a mixed state, all of the singular values of $\sqrt{\rho}J_i\sqrt{\rho}$ must vanish for each $i$, so it must be identically zero. This means all of its matrix elements in $\rho$'s eigenbasis must vanish, so we must have $\rho=\sum_k q_k |\psi_k\rangle\langle \psi_k|$ with orthonormal $\{|\psi_k\rangle\}$ where $\langle \psi_k|J_i|\psi_{k^\prime}\rangle=0$ for all $k$, $k^\prime$, and $i$.

\section{Connections and implications}
As mentioned in {Eq.~\eqref{SCSpure}}, for pure states the spin coherence scale can be written as {a sum of variances.}
Because variances so often appear in uncertainty relations and sensing applications, we are rewarded with connections between the spin coherence scale and other quantities in quantum information. We proceed by explaining how our measure relates to quantum metrology and rotation sensing, the loss of a state's purity with isotropic noise, and properties of quasiprobability distributions such as their temporal evolutions.

\subsection{Metrology, rotation sensing}
An important task in metrology is sensing the parameters of a rotation $R(\theta,\mathbf{n})$. These arise in physical situations ranging from determining inertial frames for gyroscopes~\cite{KolenderskiDemkowiczDobrzanski2008} to magnetometry~\cite{Houetal2020} and polarimetry~\cite{Pedrametal2024}. Fundamentally, rotation sensing is more complicated in quantum theory than phase estimation because one either has to take into account the possibility of different rotation axes or to explicitly measure the two angular parameters of the rotation axis. The consideration of mixed states for quantum sensing is not conducted here because mixed states are inferior to their pure counterparts for metrological tasks \cite{SidhuKok2020}.

The figure of merit in (multiparameter) quantum metrology is the quantum Fisher information (matrix), which provides a lower bound to the precision with which each of the parameters can be estimated. This is given by
\eq{
\boldsymbol{\mathsf{Q}}=\boldsymbol{\mathsf{G}}(\theta,\mathbf{n})^\top\Cov_\psi(\mathbf{J}) \boldsymbol{\mathsf{G}}(\theta,\mathbf{n}),
} where $\boldsymbol{\mathsf{G}}(\theta,\mathbf{n})$ is a $3\times 3$ real orthogonal matrix that accounts for the coordinate system in which you want to estimate your parameters (for example, replacing the angle and axis by three Euler angles), $^\top$ denotes the matrix transpose, and $\Cov_\psi(\mathbf{J})$ is the sensitivity covariance matrix with elements \cite{Goldbergetal2021rotationspublished}
\eq{
[\Cov_\psi(\mathbf{J})]_{i,j}=\frac{\left\langle J_iJ_j+J_jJ_i\right\rangle_\psi}{2}-\langle J_i\rangle_\psi\langle J_j\rangle_\psi.
} Then, the multiparameter quantum Cram\'er-Rao bound \cite{Paris2009} dictates that the covariance matrix of estimating any trio of parameters is lower bounded by $\boldsymbol{\mathsf{Q}}^{-1}$. Or, if the axis is known and all one wants to estimate is the angle, a similar expression leads to \cite{SidhuKok2020} \eq{\Delta^2\theta\geq \frac{1}{Q_{\theta\theta}}=\frac{1}{4\Var_\psi(\mathbf{J}\cdot\mathbf{n})},
\label{eq:single parameter CRB}} due to $\mathbf{J}\cdot\mathbf{n}$ being the generator responsible for rotations by $\theta$ as in Eq.~\eqref{eq:rot op}. The ultimate goal is then to find states that maximize $\Var_\psi(\mathbf{J}\cdot\mathbf{n})$ or maximize $\boldsymbol{\mathsf{Q}}$ or maximize $\Cov_\psi(\mathbf{J})$ in some sense.

Consider estimating the rotation angle $\theta$ for a variety of known rotation axes $\mathbf{n}$. If we average over all rotation axes the quantum Fisher information that a single state $|\psi\rangle$ has about that $\theta$, we find
\eq{
\frac{1}{4\pi}\int_0^{2\pi}d\Phi\int_0^\pi\sin\Theta d\Theta Q_{\theta\theta}&=\frac{4}{3}\sum_{i=1}^3\Var_\psi(J_i)\\
&=\frac{4J}{3}\SCS^2(|\psi\rangle).
} Pure states with increased spin coherence scales lead to better abilities to isotropically sense rotation angles, while increased spin also helps as a resource. If we instead average Eq.~\eqref{eq:single parameter CRB} over all rotation axes, we can use Jensen's inequality to find
\eq{
\frac{1}{4\pi}\int_0^{2\pi}d\Phi\int_0^\pi\sin\Theta d\Theta \frac{1}{Q_{\theta\theta}} \geq\frac{3}{4J\SCS^2(|\psi\rangle)},
} again providing the most sensitivity and thus the lowest axis-averaged $\Delta^2\theta$ for states that maximize the spin coherence scale, with Jensen's inequality requiring some extra state properties for saturation~\cite{Goldbergetal2020extremal}.

Then consider estimating all three parameters of the rotation. Since there are tradeoffs in the sensitivities for each parameter and one can choose a variety of coordinate systems, the first tool often used is to maximize 
\eq{
\Tr(\Cov_\psi(\mathbf{J}))=J\SCS^2(|\psi\rangle),
} which is the only part of the quantum Fisher information matrix that depends on the state and not on the parametrization. It is clear that the spin coherence scale directly dictates how good a spin-$J$ state is for simultaneously estimating all three rotation parameters. This may be formalized by choosing the intrinsic weight matrix for how to properly account for all three parameters of a rotation based on the metric tensor of $\mathfrak{su}$(2), which leads to the weighted mean squared error in estimating all three rotation parameters being lower bounded by $\Tr[\Cov_\psi(\mathbf{J})^{-1}]$~\cite{Goldbergetal2021intrinsic}. Jensen's inequality again dictates that the lower bound is $1/\Tr(\Cov_\psi(\mathbf{J}))$\sugg{, as plotted in Ref.~\cite{Goldbergetal2021rotationspublished} for various states}, so we again find the direct connection that increasing the spin coherence scale of a pure state makes it more useful for rotation sensing.

\subsection{Noise susceptibility}
The spin coherence scale captures how a state undergoing the depolarization channel of Eq.~\eqref{eq:depol Lindblad} has its purity change over time. In the analogous case of a continuous-variable state losing photons and having its QCS encapsulate the purity evolution, we proved some powerful theorems for the convexity and monotonicity of purity versus time~\cite{LupuGladsteinetal2025}. We can apply similar techniques here to find similar results, with slightly easier calculations due to spins' steady state being maximally mixed. An alternate derivation of these results using Ref.~\cite{RivasLuis2013}'s analytic solution of the time evolution will be presented afterward. The upshot is that purity is completely monotonic and log-convex with noise such that the spin coherence scale is also monotonic and convex as it evolves via the noise channel.

We presently show that a state's purity always decreases monotonically and convexly with time under Eq.~\eqref{eq:depol Lindblad}. For the monotonicity, we rewrite the purity as
\eq{
\mathcal{P}(\rho)=Q_2(\rho||\mathbb{I})
} for the 2-quasi-relative entropy $Q_2(A||B)=\Tr(A^2 B^{-1})$ with the support of $A$ contained in the support of $B$~\cite{MosonyiHiai2011}. Then, since $\mathbb{I}$ is a steady state of the Lindbladian evolution and $\rho\to\rho(t)$ under Eq.~\eqref{eq:depol Lindblad}, Ref.~\cite{MosonyiHiai2011}'s guarantee that $Q_2(\mathcal{E}(A)||\mathcal{E}(B))\leq Q_2(A||B)$ for completely positive, trace-preserving maps implies that 
\eq{
\mathcal{P}(\rho(t)){\geq} \mathcal{P}(\rho(t^\prime))\quad\forall t\leq t^\prime.
\label{eq:purityDecreasing}}


To confirm this property another way, we introduce the machinery used in Ref.~\cite{RivasLuis2013} that expands any spin-$J$ state in the basis of spherical tensors 
\eq{
\rho=\sum_{K=0}^{2J}\sum_{q=-K}^K\rho_{Kq}T_{Kq},
\label{eq:rho spherical tensor}
}where
\eq{
T_{Kq}=\sqrt{\frac{2K+1}{2J+1}}\sum_{mm^\prime=-J}^J C_{Jm,Kq}^{Jm^\prime}|J,m^\prime\rangle\langle J,m|
} are the spherical tensor operators that satisfy $\Tr[T_{Kq}^{} T_{K^\prime q^\prime}^\dagger]=\delta_{KK^\prime}\delta_{qq^\prime}$ and $T_{Kq}^\dagger=(-1)^q T_{K-q}$, and $C_{Jm,Kq}^{Jm^\prime}$ are Clebsch-Gordan coefficients. The state multipoles are thus $\rho_{Kq}=~\Tr(\rho T_{Kq}^\dagger)=(-1)^q\rho^*_{K-q}$. Because the spherical tensor operators transform covariantly as a tensor under SU(2) operations, so do the multipole moments. These multipole moments are also essential to expressing a spin state as a quasiprobability distribution on the sphere, for example using the standard spherical harmonics $Y_{Kq}(\Omega)$ to write the Wigner quasiprobability distribution~\cite{Agarwal1981,VarillyGraciaBondia1989,Klimov2002}
\eq{
W(\Omega)=\sqrt{\frac{4\pi}{2J+1}}\sum_{Kq}\rho_{Kq}Y_{Kq}(\Omega).
\label{eq:Wigner SU2}
} 
Under the time evolution of Eq.~\eqref{eq:depol Lindblad}, the multipole moments simply decay as 
\eq{\rho_{Kq}(t)=\rho_{Kq}(0)\eu^{-K(K+1)t/2J},\label{eq:rhoKq with t}} such that the purity evolves as
\eq{
\mathcal{P}(\rho{(t)})=\sum_{Kq}|\rho_{Kq}(0)|^2\eu^{-K(K+1)t/J}. \label{eq: derivative purity}
} Taking derivatives and noting that each term in the sum always shares the same sign, we immediately see that 
\eq{
\frac{(-1)^n\partial^n \mathcal{P}(\rho)}{\partial t^n}\geq 0
} such that not only does purity decrease monotonically convexly with time but, moreover, that it is a completely monotonic function of time for all initial states. {Looking at Eq.~\eqref{eq:SCSasderivativePurity},} this immediately guarantees the positivity of the spin coherence scale and also dictates that it, in turn, has the monotonicity property 
\eq{
\frac{\partial}{\partial t} \left(\mathcal{P}(\rho)\SCS^2(\rho)\right)\leq 0.
} 

To prove that the spin coherence scale is also monotonic with time when evolving via Eq.~\eqref{eq:depol Lindblad}, we require log convexity of purity under the same evolution. This is still an open question for the analogous scenario in the QCS \cite{CompanionLongArxiv}, but is easier to prove here because convex combinations of exponential decays or growths are always log convex. To see, this, take any positive coefficients $\lambda_K$ and real factors $f_K$ with the same sign in a sum $\Lambda(t)=\sum_K \lambda_K \exp(-f_K t)$. Taking the logarithm and then the time derivative, we find
$\partial\ln\Lambda(t)/\partial t=\sum_K (-f_K)\lambda_K\exp(-f_Kt)/\Lambda(T)$. {Choosing $\Lambda(t)$ to be the purity $\mathcal{P}(\rho(t))$ in} Eq.~\eqref{eq:SCSasderivativePurity}, this implies that the spin coherence scale is always positive. Then, inspecting the usual second derivative $\partial^2 \ln \Lambda(t)/\partial t^2=(\Lambda(t)\partial^2 \Lambda(t)/\partial t^2-(\partial \Lambda(t)/\partial t)^2)/(\Lambda(t))^2$, where here
\eq{
\Lambda(t)\frac{\partial^2 \Lambda(t)}{\partial t^2}-\left(\frac{\partial \Lambda(t)}{\partial t}\right)^2=&\sum_{KL}\lambda_K\lambda_L \eu^{-(f_K +f_L)t} f_K(f_K-f_L)\\
=&\sum_{K<L}\lambda_K\lambda_L \eu^{-(f_K +f_L)t} (f_K-f_L)^2
} is positive for all $t$, we conclude that the spin coherence scale decreases monotonically with $t$ {under the depolarization channel}.


\subsection{Quasiprobability distributions}
In the continuous-variable case, many properties of the QCS were derived thanks to the fact that the purity can be written as an integral of $s$- and $(-s)$-ordered quasiprobability distributions over the whole phase space \cite{CompanionLongArxiv}; c.f. Eq.~\eqref{eq:purity from quasi CV}. In the spin setting, we will now introduce some background machinery to understand the slightly less famous $s$-ordered quasiprobability distributions for SU(2). We will see that, in this setting too, the purity can be computed as a similar integral, so many connections to $\SCS^2$ that rely on properties of purity will hold.

The $s$-ordered quasiprobability distributions for spins generalize the Wigner distribution in Eq.~\eqref{eq:Wigner SU2} to \cite{Klimov2002}
\eq{
W^{(s)}_\rho(\Omega)=\sqrt{\frac{4\pi}{2J+1}}\sum_{Kq}\left(C_{JJ,K0}^{JJ}\right)^{-s}\rho_{Kq}Y_{Kq}(\Omega).
\label{eq:s ordered SU2}
} As is clear, the Clebsch-Gordan coefficients are responsible for the transition between the Husimi function
\eq{
W^{(-1)}_\rho(\Omega)=\langle\Omega^{(J)}|\rho|\Omega^{(J)}\rangle
} and the Glauber-Sudarshan-type function $W^{(1)}(\Omega)$ that furnishes the diagonal representation
\eq{
\rho=\frac{2J+1}{4\pi}\int d\Omega W^{(1)}_\rho(\Omega)|\Omega^{(J)}\rangle\langle\Omega^{(J)}|,
}
where $d\Omega=\sin\theta d\theta d\phi$ is the invariant measure on the \sugg{2-}sphere \sugg{$S^2=\{(\theta,\phi)\, |\, 0\leq \theta\leq \pi,0\leq \phi<2\pi\}$} and the $(2J+1)/4\pi$ factors that appear throughout are familiar from spin-coherent states' resolution of identity $\mathbb{I}=\frac{2J+1}{4\pi}\int d\Omega |\Omega^{(J)}\rangle\langle\Omega^{(J)}|$. 
As with the Heisenberg-Weyl group, the Husimi function ($s=-1$) is positive everywhere {for all states} and classical states have {positive} 
$P$ functions ($s=1$) everywhere; however, the Wigner functions for spin-coherent states must be negative somewhere and thus Wigner negativity is a complicated quantifier of quantumness~\cite{Davisetal2021} \sugg{that we study in Sec.~\ref{sec:wig neg}.}.
These informationally complete versions of a state can be extended to any operator and provide the overlap relation
\eq{
\Tr(AB)=\frac{2J+1}{4\pi}\int d\Omega W_A^{(-s)}(\Omega)W_B^{(s)}(\Omega)
\label{eq:overlap wigner SU2}
} along with the quasiprobability properties $\frac{2J+1}{4\pi}\int d\Omega W^{(s)}_\rho(\Omega)=1$ and $W^{(s)}_\rho(\Omega)^*=W^{(s)}_\rho(\Omega)$.

Relations between the spin coherence scale and purity dictate relations for quasiprobability distributions. First, we substitute $A=B=\rho$ into Eq.~\eqref{eq:overlap wigner SU2} and find this overlap integral to be completely monotonic with time under the depolarization channel, for any $s$. 
Then, noting that $\Tr(\rho[J_i,[J_i,\rho])=\Tr([\rho,J_i][J_i,\rho])$ and using Eqs.~\eqref{eq:overlap wigner SU2} and \eqref{eq:rhoKq with t}, Eq.~\eqref{eq:SCS commutators} can be written as: 
\eq{
\SCS^2&=\frac{\int d\Omega W^{(-s)}_\rho(\Omega) W_{\sum_{i=1}^3 [J_i,[J_i,\rho]]}^{(s)}(\Omega)}{2J\int d\Omega W^{(-s)}_\rho(\Omega) W_{\rho}^{(s)}(\Omega)}\\
&=-\frac{\int d\Omega W^{(-s)}_\rho (\Omega)W_{\partial\rho/\partial t}^{(s)}(\Omega)}{\int d\Omega W^{(-s)}_\rho (\Omega)W_{\rho}^{(s)}(\Omega)}\\
&\sugg{=
-\frac{\int d\Omega W^{(-s)}_\rho (\Omega)\Delta W_{\rho}^{(s)}(\Omega)}{2J\int d\Omega W^{(-s)}_\rho (\Omega)W_{\rho}^{(s)}(\Omega)}
}\\
&\sugg{=
\frac{\int d\Omega \nabla W^{(-s)}_\rho (\Omega)\nabla W_{\rho}^{(s)}(\Omega)}{2J\int d\Omega W^{(-s)}_\rho (\Omega)W_{\rho}^{(s)}(\Omega)}
}
,\label{eq:SCS fraction of Wigner int}
}
where we used Eq.~\eqref{eq:depol Lindblad} in the second line\sugg{, $\Delta Y_{Kq}(\Omega)=-K(K+1)Y_{Kq}(\Omega)$ for the  spherical Laplace-Beltrami operator (the angular part of the Laplacian) $\Delta$ in the third, and divergence theorem in the fourth}. 
This has a form similar to the QCS's in Eq.~\eqref{eq:QCS quasi}, where the numerator has a divergence squared or a Laplacian acting on the Wigner function. That was shown~\cite{CompanionLongArxiv} to originate from the special relationship between loss and quasiprobability distributions for position and momentum, where losing a fraction $1-\eta$ of the photons enacts $W_\rho^{(s)}(\alpha)\to W_{\rho_\eta}^{(s)}=\frac1{\eta}W_{\rho}^{(1+(s-1)/\eta)}(\alpha/\sqrt{\eta})$~\cite{RadimFilip}. Does such a relationship exist here; is there an SU(2)-invariant evolution that evolves a state's quasiprobability distribution on the sphere to one with smaller $s$?

The answer is yes in the limit of large spin $J$, which in some contexts is considered a classical limit and in all contexts is the limit where the spherical manifold of SU(2) begins to look locally flat and contracts to the Heisenberg-Weyl group. To see this, we expand the relevant Clebsch-Gordan coefficients for large $J$ to find
\eq{
\ln C_{JJ,K0}^{JJ}=\frac{1}{2}\ln\frac{\binom{4J+1}{2J-K}}{\binom{4J+1}{2J}}=-\frac
{K(K+1)}{4J}+\mathcal{O}\left(\frac{1}{J^2}\right).
} 
Using Eq.~\eqref{eq:rhoKq with t} in Eq.~\eqref{eq:s ordered SU2} and noting the parallel factors of $K(K+1)/J$  we can identify 
\eq{
W_{\rho(t)}^{(s)}(\Omega)\approx W_{\rho(0)}^{( s -2t)}(\Omega)
\label{Approx W}} to lowest order in $1/J$. Just like for Heisenberg-Weyl and loss, depolarization noise for SU(2) quasiprobability distributions monotonically lowers the order $s$; after enough time, all of the quasiprobability distributions become positive, because the $s=-1$ (Husimi) distribution is manifestly positive.

From this we find inequalities of the same style as for the QCS. For example, using Eqs.~\eqref{eq:purityDecreasing},~\eqref{eq: derivative purity},~and~{\eqref{Approx W}} 
\eq{
&\int \!\!d\Omega W^{(s-2t)}_{{\rho(0)}} (\Omega) W_{\rho (0)}^{(-s+2t)}(\Omega) \\ 
&\qquad \geq \!\!\int\!\! d\Omega W^{(s-2t^\prime)}_{{\rho(0)}}  (\Omega) W_{{\rho(0)} }^{(-s+2t^\prime)}(\Omega) 
} for all $t\leq t^\prime$, where $\rho(0)$ can be any state. We can also now discuss the spin coherence scale in terms of differentiation with respect to the ordering parameter $s$ due to $W^{(s)}_{\partial \rho(t)/\partial t}(\Omega)=\partial W^{(s)}_{ \rho(t)}(\Omega)/\partial t{\approx}-2\partial W^{(s)}_{ \rho(t)}(\Omega)/\partial s$ (for large $J$).
 Then, Eq.~\eqref{eq:SCS fraction of Wigner int} becomes, again for large $J$:
\eq{
\SCS^2\approx 2\frac{\int d\Omega W^{(-s)}_\rho (\Omega)\partial W_{\rho}^{(s)}(\Omega)/\partial s}{\int d\Omega W^{(-s)}_\rho(\Omega) W_{\rho}^{(s)}(\Omega)}.
}

The above approximations held to leading order in $1/J$; is there a time evolution that exactly leads to an evolution between quasiprobability distributions, or at least can do so to next leading order? Expanding the Clebsch-Gordan coefficient again, we find $C_{JJ,K0}^{JJ}\approx \exp(-K(K+1)(2J-1)/8J^2)$, so it looks as though the time parameter must simply be adjusted as \eq{W_{\rho(t)}^{(s)}(\Omega)\approx W_{\rho(0)}^{\left( s -\tfrac{2tJ}{J-1/2}\right)}(\Omega)} and the inequalities adjusted accordingly. The next orders after that seem to all involve polynomials in $K(K+1)$, which can be achieved by Lindblad operators of the form $L_{\mathbf{k}}\propto J_i \cdots J_j$ for the Markovian master equations $\tfrac{\partial \rho}{\partial t}=~\sum_{\mathbf{k}} \left(L_{\mathbf{k}}^{}\rho L_{\mathbf{k}}^\dagger-\frac{L_{\mathbf{k}}^\dagger L_{\mathbf{k}}^{}\rho+\rho L_{\mathbf{k}}^\dagger L^{}_{\mathbf{k}}}{2}\right)$. However, there are two caveats. First, the polynomial's coefficients must be shown to be positive. And second, $C_{JJ,K0}^{JJ}$ is not a polynomial in just $K(K+1)$: inspecting the $K$-dependent denominator, the factorials multiply to
\eq{
(2J-K)!(2J+K+1)!=(2J(2J+1)-K(K+1))\\
\times ((2J-1)2J-K(K+1))\cdots\\
\times((2J-n)(2J+1-n)-K(K+1))\cdots
\times (2K+1)!.
} Each pair $(2J-K-n)(2J+K+1-n)$ multiplies to a polynomial in $K(K+1)$, but the extra unpaired factors in $(2K+1)!$ are not such a polynomial. We leave this as an open problem in the theory of spherical tensors for achieving $\sum_{\mathbf{k}} \left(L_{\mathbf{k}}^{}T_{Kq}^{} L_{\mathbf{k}}^\dagger-\frac{L_{\mathbf{k}}^\dagger L^{}_{\mathbf{k}}T^{}_{Kq}+T^{}_{Kq} L_{\mathbf{k}}^\dagger L^{}_{\mathbf{k}}}{2}\right)\underset{?}{=}T_{Kq}\frac{\partial f(t)}{\partial t}\ln C_{JJ,K0}^{JJ}$: is there a Lindblad evolution (Markovian master equation, completely positive trace preserving evolution) that evolves SU(2) quasiprobability distributions into other SU(2) quasiprobability distributions with different orders $s$ as $W_{\rho(t)}^{(s)}(\Omega)\approx W_{\rho(0)}^{( s -f(t))}(\Omega)$?


\subsection{Relation to Wigner Negativity}
\label{sec:wig neg}

\sugg{In this section, we derive a sufficient condition, expressed in terms of the spin coherence scale \(\SCS^2(\rho)\), that guarantees positivity of the spin Wigner function \(W_\rho^{(0)}(\Omega)\) for all \(\Omega \in S^2\). The task of finding such states is challenging even for bosonic systems~\cite{VanHerstraetenetal2026} and in general is essential for knowing the usefulness of spin systems for quantum information tasks~\cite{Howardetal2014,Delfosseetal2017}}

We begin by taking the time derivative of Eq.~\eqref{eq: derivative purity} and substituting it into Eq.~\eqref{eq:SCSasderivativePurity} so that the spin coherence scale takes the
compact form
\begin{equation}
\SCS^2
=
\frac{1}{2J\mathcal{P}(\rho)}
\sum_{K=1}^{2J} \sum_{q=-K}^{K}
K(K+1)\,|\rho_{K q}|^2.
\label{eq:SCS-final}
\end{equation}

\begin{purpleblock}
Now, in order to establish a bound on $\SCS^2$, we decompose the Wigner function into a constant term $\overline{W} =\sqrt{\frac{4\pi}{2J+1}}\rho_{00}Y_{00}(\Omega)= 1/(2J+1)$ and the residual, angularly varying portion:
\begin{equation}
\delta W_\rho(\Omega)
:= W^{(0)}_\rho(\Omega) - \overline{W},
\end{equation}
so that only modes with \(K\ge1\) contribute. 
We now derive an explicit pointwise bound on the angular variation
\(\delta W_\rho(\Omega)\).
Using the spherical-harmonic expansion Eq. ~\eqref{eq:Wigner SU2},
\[
\delta W_\rho(\Omega)
=
\sqrt{\frac{4\pi}{2J+1}}
\sum_{K=1}^{2J}\sum_{q=-K}^{K}
\rho_{Kq}\,Y_{Kq}(\Omega),
\]
the triangle inequality gives
\begin{equation}
|\delta W_\rho(\Omega)|
\le
\sqrt{\frac{4\pi}{2J+1}}
\sum_{K=1}^{2J}\sum_{q=-K}^{K}
|\rho_{Kq}|\,|Y_{Kq}(\Omega)|.
\label{eq:triangle}
\end{equation}

\noindent
Inserting the factor \(\sqrt{K(K+1)/K(K+1)}\) and applying the Cauchy--Schwarz inequality
over the magnetic index \(q\):
\begin{align}
\sum_{q=-K}^{K} |\rho_{Kq}|\,|Y_{Kq}|
&=
\sum_{q=-K}^{K}
\frac{\sqrt{K(K+1)}\,|\rho_{Kq}|}{\sqrt{K(K+1)}}\,|Y_{Kq}|
\nonumber\\
&\le
\sqrt{\sum_{q=-K}^{K} K(K+1)\,|\rho_{Kq}|^2}\;
\sqrt{\sum_{q=-K}^{K} \frac{|Y_{Kq}(\Omega)|^2}{K(K+1)}}.
\label{eq:CS-q}
\end{align}

\noindent
The second factor is evaluated using the spherical addition theorem
\cite[Eq.~(5.10.1)]{Varshalovichetal1988},
\begin{equation}
\sum_{q=-K}^{K} |Y_{Kq}(\Omega)|^2
=
\frac{2K+1}{4\pi},
\end{equation}
which yields
\begin{equation}
\sqrt{\sum_{q=-K}^{K} \frac{|Y_{Kq}(\Omega)|^2}{K(K+1)}}
=
\sqrt{\frac{2K+1}{4\pi\,K(K+1)}}.
\label{eq:addition}
\end{equation}

\noindent
Substituting Eqs.~\eqref{eq:CS-q} and \eqref{eq:addition} into
Eq.~\eqref{eq:triangle}, we obtain
\begin{align}
|\delta W_\rho(\Omega)|
&\le
\sqrt{\frac{4\pi}{2J+1}}
\sum_{K=1}^{2J}
\sqrt{\sum_{q=-K}^{K} K(K+1)\,|\rho_{Kq}|^2}
\\
&\quad\times
\sqrt{
\frac{2K+1}{4\pi\,K(K+1)}
}.
\label{eq:sum-K}
\end{align}

\noindent
Applying the Cauchy--Schwarz inequality once more, now over the index \(K\),
gives
\begin{align}
|\delta W_\rho(\Omega)|
&\le
\sqrt{\frac{4\pi}{2J+1}}
\sqrt{
\sum_{K=1}^{2J}\sum_{q=-K}^{K} K(K+1)\,|\rho_{Kq}|^2
}
\\
&\quad\times
\sqrt{
\sum_{K=1}^{2J} \frac{2K+1}{4\pi\,K(K+1)}
}.
\label{eq:CS-K}
\end{align}

\noindent
The first sum is proportional to the spin coherence scale,
\[
\sum_{K=1}^{2J}\sum_{q=-K}^{K} K(K+1)\,|\rho_{Kq}|^2
=
2J\,\mathcal{P}\,\SCS^2,
\]
The second sum can be evaluated explicitly in terms of harmonic numbers
\(H_n=\sum_{k=1}^n k^{-1}\),
\begin{equation}
\sum_{K=1}^{2J} \frac{2K+1}{4\pi\,K(K+1)}
=
\frac{1}{4\pi}\big(H_{2J}+H_{2J+1}-1\big).
\end{equation}

\noindent
Combining these results yields the pointwise bound
\begin{equation}
\sup_{\Omega} |\delta W_\rho(\Omega)|
\le
\sqrt{
\frac{2J\,\mathcal{P}\,\SCS^2}{2J+1}
\;
\big(H_{2J}+H_{2J+1}-1\big)
}.
\label{eq:deltaW-bound}
\end{equation}

Requiring \(|\delta W_\rho(\Omega)| \le \overline{W}\) for all \(\Omega\) yields a sufficient condition for Wigner positivity:
\begin{equation}
\SCS^2
\le
\frac{1}
{2J(2J+1)\,\mathcal{P}\,
\big(H_{2J}+H_{2J+1}-1\big)}
\;\Longrightarrow\;
W^{(0)}_\rho(\Omega) \ge 0 \;\; \forall\,\Omega .
\end{equation}

This bound shows that a sufficiently small spin coherence scale value of a given state precludes the appearance of Wigner negativity.

\begin{figure}[t]
\centering

\begin{minipage}[t]{0.48\columnwidth}
    \centering
    \includegraphics[width=\linewidth]{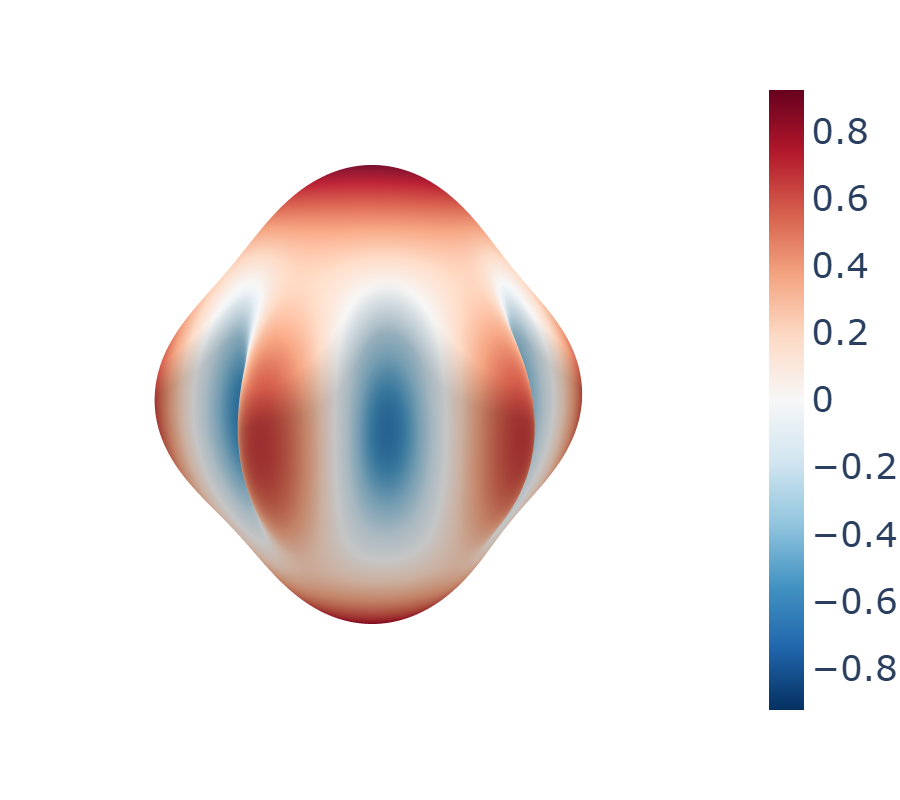}
    \vspace{0.3em}
    
    {\small (a) Wigner function of the NOON state for $J=3$.}
\end{minipage}
\hfill
\begin{minipage}[t]{0.48\columnwidth}
    \centering
    \includegraphics[width=\linewidth]{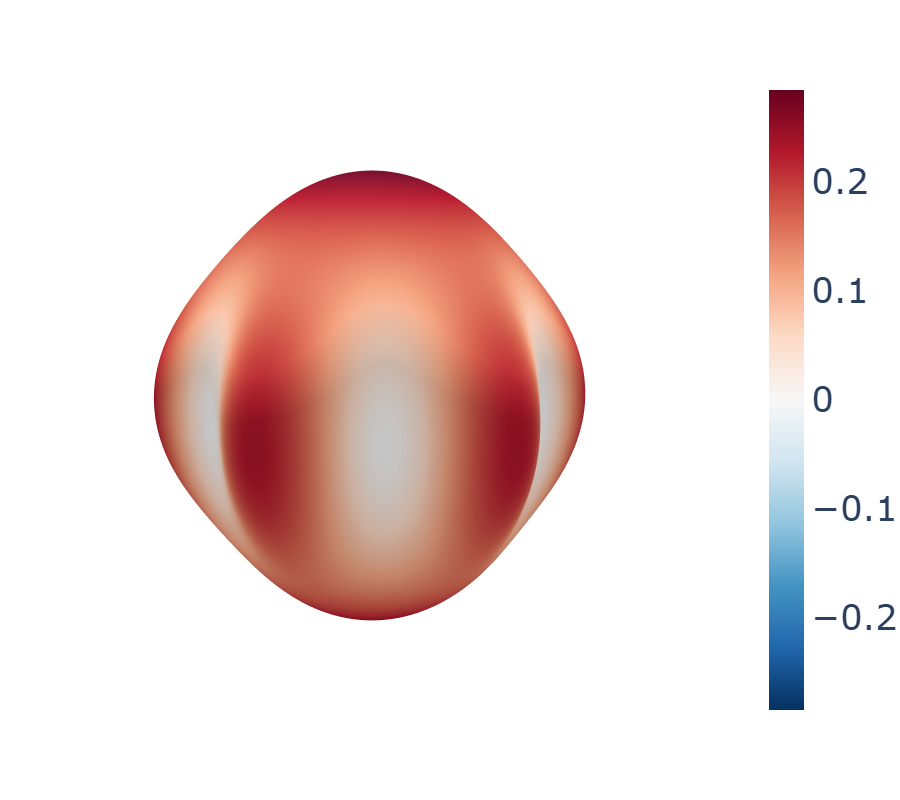}
    \vspace{0.3em}
    
    {\small (b) NOON state mixed with the maximally mixed state for $J=3$, with mixing parameter $p \approx 0.18$.}
\end{minipage}

\caption{Wigner functions visualized as local radial deformations of the sphere. The pure NOON state exhibits pronounced oscillatory structure and negativity, while sufficient mixing with the maximally mixed state suppresses fine-scale features and removes negativity.}
\label{fig:noon_wigner_comparison}
\end{figure}

\sugg{We plot this bound for some families of states $\rho_p=p|\psi\rangle\langle\psi|+(1-p)\mathds{1}/(2J+1)$; representative Wigner functions can be seen in Fig.~\ref{fig:noon_wigner_comparison}. For each such state, the Wigner function will be negative somewhere for $p=1$ (pure states) and be positive everywhere for $p=0$ (maximally mixed states). The purities of these states can be computed analytically as $\mathcal{P}(\rho_p)=p^2+(1-p^2)/(2J+1)$ and the double-commutator portions also possess a simple scaling relation $\operatorname{Tr}\!\left( \sum_i \rho_p [J_i,[J_i,\rho_p]] \right) = p^2 \operatorname{Tr}\!\left( \sum_i \rho_1 [J_i,[J_i,\rho_1]] \right)$. The full families are plotted in Fig.~\ref{fig:scs depol}, for spin coherent, NOON, Berry-Wiseman, and Yurke states. Here, the Berry-Wiseman states are the spin states (or equivalently two photon states with fixed total photon number) for optimal phase precision with  single shot estimation \cite{PhysRevLett.85.5098}, and are defined as:
\begin{equation}
|\psi_{\mathrm{BW}}\rangle
=
\frac{1}{\sqrt{J+1}}
\sum_{m=-J}^{J}
\sin\!\left(
\frac{(J + m + 1)\pi}{2J + 2}
\right)
\,|J,m\rangle.
\end{equation}}
The Yurke states are spin-squeezed states that also find application in interferometric phase estimation  \cite{Combes_2004}, and for integer spin \(J\) are given by
\begin{equation}
|\psi_{\mathrm{Yurke}}\rangle
=
\frac{\sin\alpha}{\sqrt{2}}\,|J,1\rangle_2
+ \cos\alpha\,|J,0\rangle_2
+ \frac{\sin\alpha}{\sqrt{2}}\,|J,-1\rangle_2,
\end{equation}
where \(|J,m\rangle_2\) are eigenstates of \(J_2\); in the limit \(\alpha \to 0\), they achieve squeezing scaling as \(1/J\). In Fig. \ref{fig:scs depol}, $\alpha = 0.15$.

\begin{figure*}
    \centering
    \includegraphics[width=\textwidth]{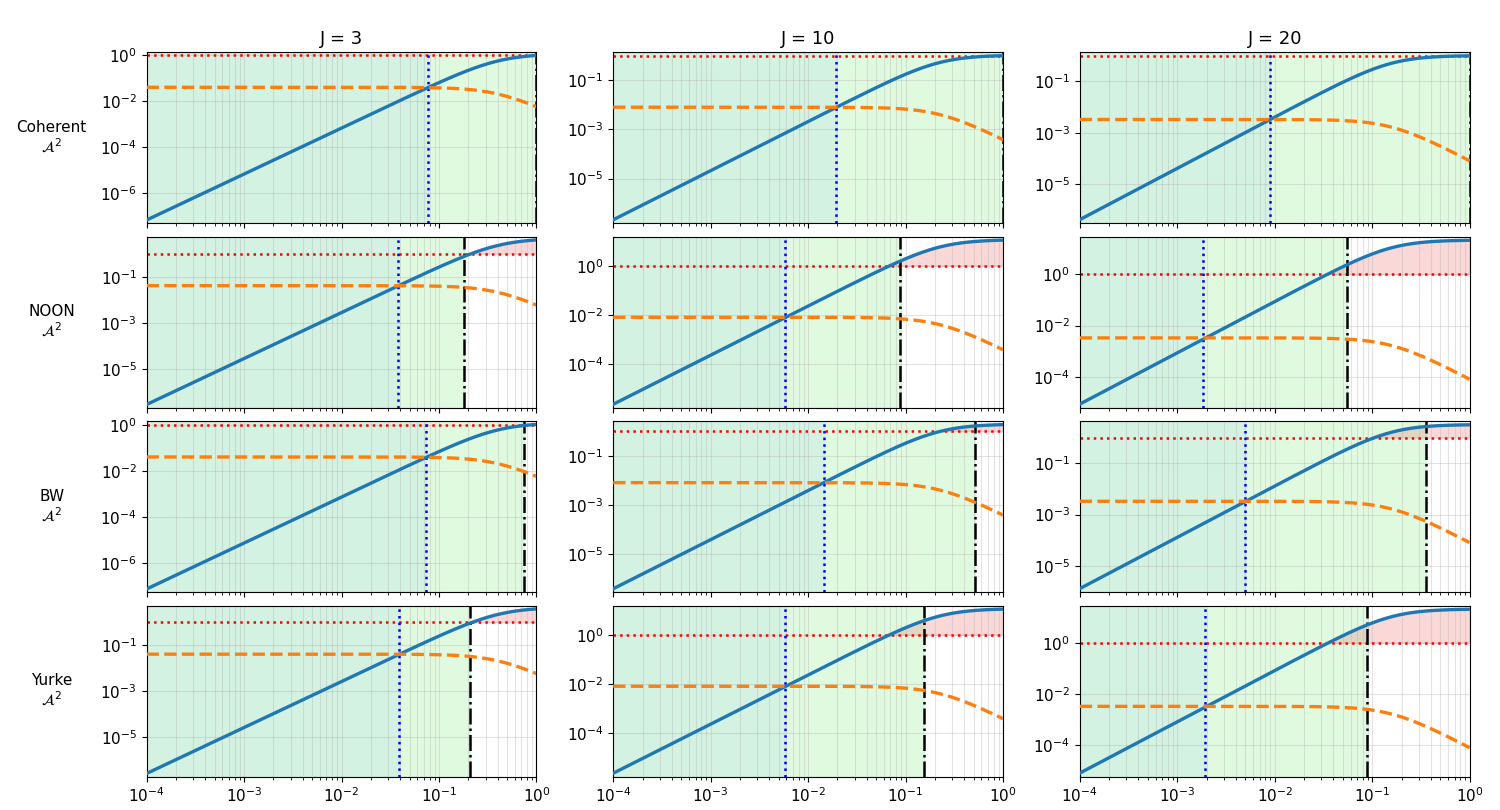}
    \caption{
    Spin coherence scales $\SCS^2$ for families of states (rows) in different dimensions (columns) and with different mixing parameters $p$. The scales decrease monotonically from pure ($p=1$) to maximally mixed ($p=0$) states. The blue solid curve shows $\SCS^2$, the orange dashed line the analytic positivity bound, and the red dotted line the classicality threshold $\SCS^2 = 1$. Three regimes are highlighted: (i) states with $\SCS^2 > 1$ (red shaded region) are certified to be nonclassical; (ii) the vertical blue dotted line marks $p_{\mathrm{pos}}$, below which the positivity bound guarantees a nonnegative Wigner function (dark green region); (iii) the vertical black dash-dotted line marks the true critical value $p_{\mathrm{crit}}$, below which the Wigner function is actually nonnegative. The light green region between $p_{\mathrm{pos}}$ and $p_{\mathrm{crit}}$ indicates the gap between the bound and the exact transition. The tightness of the bound is governed by how nearly the triangle and Cauchy--Schwarz inequalities are saturated. Values of $p_{\mathrm{pos}}$ and $p_{\mathrm{crit}}$ can be found in Appendix \ref{app:pos thresholds}.}
    \label{fig:scs depol}
\end{figure*}

\subsection{Structure of Bound-Saturating States}
\label{sec:extremal_states}

We now characterize the states that saturate the bound on the Wigner fluctuation. Since the bound is derived through successive applications of the triangle inequality and Cauchy--Schwarz inequalities, saturation requires equality at each step. Rather than repeating the full derivation, we summarize the resulting constraints.

Equality in Eq.~\eqref{eq:triangle} enforces a phase-alignment condition, requiring that all contributing terms add constructively at the point $\Omega_*$ where the fluctuation is maximized. Equality in Eq.~\eqref{eq:CS-q} imposes a constraint within each fixed-$K$ sector, fixing the relative structure of the multipole components $\rho_{Kq}$. Finally, Eq.~\eqref{eq:CS-K} enforces a global constraint across different $K$, determining their relative weights. Taken together, these conditions uniquely determine the extremal multipole form,
\begin{equation}
\rho_{Kq}
=
\lambda \frac{Y_{Kq}^*(\Omega_*)}{K(K+1)},
\end{equation}
for some real parameter $\lambda$.

It is convenient to express the corresponding operator in a normalized form consistent with the Wigner expansion,
\begin{equation}
\rho(\lambda)
=
\frac{\mathds{1}}{2J+1}
+
\lambda
\sum_{K=1}^{2J}
\frac{\sqrt{(2K+1)/(2J+1)}}{K(K+1)}\,T_{K0},
\end{equation}
where we have chosen a representative aligned along a fixed axis. Hermiticity and unit trace are automatic for real $\lambda$, and physicality is therefore determined entirely by positivity, which restricts $\lambda$ to a finite interval. Within this interval, all such states saturate the fluctuation bound.

One can show that any state of this form is axisymmetric up to rotation, i.e.\ all extremal states are related by rotations to a representative supported entirely on the $q=0$ sector. We defer the proof of this statement to Appendix~\ref{app:axisymmetry}.

To verify that the extremal family contains physical states, we evaluate the spectrum of $\rho(\lambda)$ for a representative case. For $J=10$, we determine the value of $\lambda$ for which
\begin{equation}
\sup_{\Omega} \delta W(\Omega) = -\frac{1}{2J+1},
\end{equation}
and confirm that the corresponding operator remains positive (smallest eigenvalue $\approx 0.02$). The associated Wigner function is shown in Fig.~\ref{fig:extremal_wigner}, where the maximal fluctuation is attained while maintaining nonnegativity.

\begin{figure}[H]
    \centering
    \includegraphics[width=0.4\textwidth]{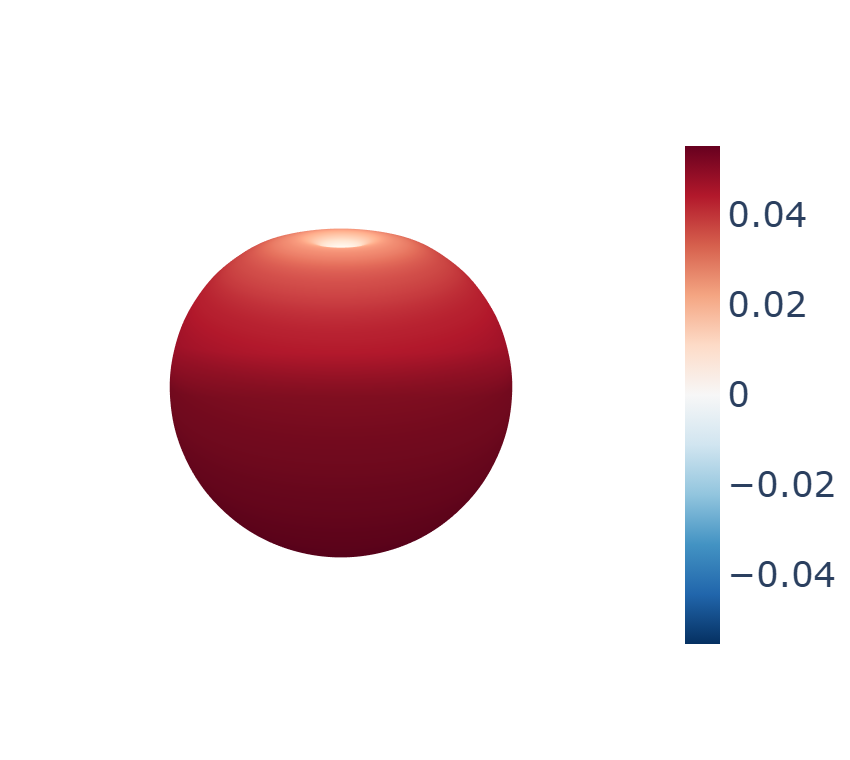}
    \caption{Wigner function of an extremal state for $J=10$. The maximal fluctuation $\sup_{\Omega} |\delta W_\rho(\Omega)| = 1/(2J+1)$ is attained while maintaining nonnegativity.}
    \label{fig:extremal_wigner}
\end{figure}

We have thus identified the full family of states saturating the fluctuation bound. Since this family is completely determined by the equality conditions of the inequalities used in the derivation, no state can exceed this bound. The derived bound is therefore the optimal state-independent bound.

\end{purpleblock}

\begin{purpleblock}

\section{Measurement Protocol for the Spin Coherence Scale}

The spin coherence scale also admits an operational interpretation in terms of collective observables acting on two identical copies of the state. In this section, we present a linear-optical two-copy protocol for measuring $\SCS^2$ for a spin-$J$ system. The protocol closely parallels the interferometric scheme developed for the QCS, and is experimentally feasible using standard linear-optical components, following an architecture similar to that of Ref.~\cite{Griffet}.

\subsection{Two-Copy Expression for the Spin Coherence Scale}

Let $\rho$ be a spin-$J$ state and consider two identical copies $\rho \otimes \rho$. Define the collective sum and difference spin operators
\begin{equation}
\mathbf{J}_s = \mathbf{J}^{(1)} + \mathbf{J}^{(2)}, \qquad
\mathbf{J}_d = \mathbf{J}^{(1)} - \mathbf{J}^{(2)}.
\end{equation}
These satisfy the identity
\begin{equation}
\mathbf{J}_s^2 + \mathbf{J}_d^2 = 4J(J+1)\,\mathbb{I}.
\end{equation}

Introducing the swap operator $\mathrm{S}$ acting on the two-copy Hilbert space, the spin coherence scale can be written compactly as
\begin{equation}
\mathcal{A}^2
=
\frac{1}{2J\,\mathcal{P}(\rho)}
\operatorname{Tr}\!\left[
(\rho \otimes \rho)\,
\mathrm{S}\,
\mathbf{J}_d^2
\right].
\label{eq:A2_twocopy}
\end{equation}
Equivalently,
\begin{equation}
\operatorname{Tr}\!\left[
(\rho \otimes \rho)\,
\mathrm{S}\,
\mathbf{J}_s^2
\right]
=
4J(J+1)\,\mathcal{P}
-
2J\,\mathcal{P}\,\mathcal{A}^2.
\label{eq:A2_Js}
\end{equation}
Thus, $\mathcal{A}^2$ is fully determined by the expectation value of a single collective two-copy observable, together with the purity.

\subsection{Jordan--Schwinger Encoding}

To realize these measurements experimentally, we employ the Jordan--Schwinger representation~\cite{Schwinger1965AngularMomentum,SakuraiNapolitano2020}, which encodes a spin-$J$ system into two bosonic modes $a$ and $b$ 
\begin{align}  
&J_1 = \tfrac{1}{2}(a^\dagger b + b^\dagger a), \qquad
J_2 = \tfrac{1}{2i}(a^\dagger b - b^\dagger a),\nonumber \\
&J_3 = \tfrac{1}{2}(a^\dagger a - b^\dagger b),
\end{align}
with the constraint $a^\dagger a + b^\dagger b = 2J$.
In this encoding, angular momentum eigenstates $|J,m\rangle$ correspond to dual-rail Fock states
\begin{equation}
|J,m\rangle \leftrightarrow |n_a = J+m,\; n_b = J-m\rangle.
\end{equation}
The two bosonic modes may be implemented as orthogonal polarization modes of a single spatial beam, or as two spatial modes with identical polarization~\cite{RevModPhys.79.135}.

\subsection{Linear-Optical Two-Copy Interference}

Two identical copies of the encoded state are prepared in mode pairs $(a^{(1)},b^{(1)})$ and $(a^{(2)},b^{(2)})$. Each pair of corresponding modes is interfered on a balanced (50{:}50) beam splitter to yield the output modes
\begin{equation}
\begin{aligned}
a_\pm &= \tfrac{1}{\sqrt{2}}\!\left(a^{(1)} \pm a^{(2)}\right),\\
b_\pm &= \tfrac{1}{\sqrt{2}}\!\left(b^{(1)} \pm b^{(2)}\right).
\end{aligned}
\end{equation}
Photon-number-resolving detectors are placed at the output ports. The parity of the total photon number in the antisymmetric outputs,
\begin{equation}
\mathrm{S} = (-1)^{n_{a_-} + n_{b_-}},
\end{equation}
directly measures the swap operator between the two copies, yielding the purity $\mathcal{P}$.

At the same time, the population imbalance at the outputs satisfies
\begin{equation}
n_{a_+} + n_{a_-} - n_{b_+} - n_{b_-}
=
2\!\left(J_3^{(1)} + J_3^{(2)}\right).
\end{equation}

Combining these observables yields
\begin{align}
&\operatorname{Tr}\!\left[(\rho\otimes\rho)\,\mathrm{S}\,(J_3^{(1)}+J_3^{(2)})^2\right] \\
&\quad\quad=
\frac{1}{4}
\Big\langle
(-1)^{n_{a_-}+n_{b_-}}
(n_{a_+}+n_{a_-}-n_{b_+}-n_{b_-})^2
\Big\rangle .
\end{align}
These transformations are schematized in Figs.~\ref{fig:schematic1}~and~\ref{fig:schematic2} for encoding in spatial and polarization modes, respectively.

\subsection{Accessing All Spin Components}

To measure collective spin observables along arbitrary directions, identical SU(2) rotations are applied locally to each copy prior to interference,
\begin{equation}
\begin{pmatrix} a \\ b \end{pmatrix}
\longrightarrow
{R(\theta,\mathbf{n})}
\begin{pmatrix} a \\ b \end{pmatrix}R(\theta,\mathbf{n})^\dagger
.
\end{equation}
Denoting by
\begin{equation}
U_R = R \otimes R
\end{equation}
the induced transformation on the two-copy Hilbert space, the measured observable becomes
\begin{equation}
U_R^\dagger\, \mathrm{S}\, (J_3^{(1)}+J_3^{(2)})^2\, U_R .
\end{equation}

The crucial point is that the swap operator is invariant under identical local rotations on the two copies:
\begin{equation}
(R\otimes R)\,\mathrm{S}\,(R^\dagger\otimes R^\dagger)=\mathrm{S}.
\label{eq:swap_rotation_invariant}
\end{equation}
Thus the interferometric part of the protocol that measures the swap observable is unchanged; only the spin component being probed is rotated.

Under the same rotation, the collective spin transforms as
\begin{equation}
(R\otimes R)^\dagger (J_3^{(1)}+J_3^{(2)})(R\otimes R)
=
J_{\mathbf{n}}^{(1)}+J_{\mathbf{n}}^{(2)},
\end{equation}
for an appropriate choice of axis $\mathbf{n}$.

It follows that after applying identical rotations to both copies, the same detection scheme measures
\begin{equation}
\operatorname{Tr}\!\left[
(\rho\otimes\rho)\,
\mathrm{S}\,
\bigl(J_{\mathbf{n}}^{(1)}+J_{\mathbf{n}}^{(2)}\bigr)^2
\right].
\label{eq:rotated_component_measurement}
\end{equation}
In polarization encoding, these rotations are implemented using wave plates; in spatial-mode encoding, using Mach-Zehnder interferometers with tunable phase shifters. Choosing $R = \mathbb{I}$,  $R(\pi/2,(0,1,0)^\top)$, and $R(-\pi/2,(1,0,0)^\top)$, 
enables measurement of the $J_3$, $J_1$, and $J_2$ components, respectively. Summing the corresponding outcomes yields the quantity in Eq.~\eqref{eq:A2_Js}, from which the spin coherence scale $\mathcal{A}^2$ is directly extracted.

\begin{figure}[t]
    \centering
    \includegraphics[width=\linewidth]{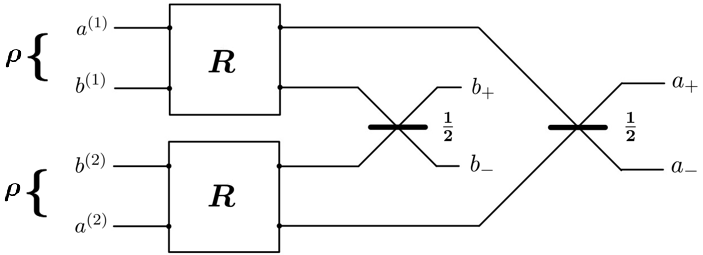}
    \caption{
    Schematic of the two-copy linear-optical measurement protocol for the spin coherence scale.
    Two identical copies of the spin-$J$ state are encoded in bosonic modes
    $(a^{(1)},b^{(1)})$ and $(a^{(2)},b^{(2)})$.
    Identical SU(2) rotations $R$ are applied locally to each copy,
    followed by interference of corresponding modes on balanced beam splitters.
    Photon-number-resolving detection allows simultaneous measurement of the swap operator
    $\mathrm{S}$ and collective spin observables.
    Choosing $R = \mathbb{I}$,  $R(\pi/2,(0,1,0)^\top)$, and $R(-\pi/2,(1,0,0)^\top)$, enables measurement of the $J_3$, $J_1$, and $J_2$ components, respectively.
    }
    \label{fig:schematic1}
\end{figure}

\begin{figure}[t]
    \centering
    \includegraphics[width=\linewidth]{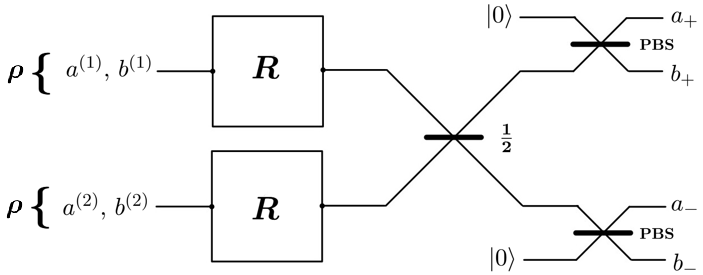}
    \caption{
    Alternative formulation for the protocol, using polarization degrees of freedom. Here, the two copies of $\rho$ are each realized using one spatial mode, and two polarization degrees of freedom (here we associate $a^{(1,2)}$ with H polarization, $b^{(1,2)}$ with V polarization). The SU(2) rotations are realized by compositions of waveplates, and polarization beamsplitters are used to separate $a_+, a_-, b_+, b_-$ after the input modes are all interfered on a balanced beamsplitter.}
    \label{fig:schematic2}
\end{figure}
\end{purpleblock}

\section{Extensions to SU($n$)}
Now that we have the coherence scales for the Heisenberg-Weyl group (the QCS) and for SU(2), what other groups can we tackle? Many of the expressions we worked out here extend naturally to SU($n$); we highlight the main results and give details of some calculations in Appendix~\ref{app:suN details}.

The Lie algebra $\mathfrak{su}(n)$ that generates the Lie group SU($n$) is spanned by $d=n^2-1$ traceless, Hermitian generators $J_i$. The SU($n$) coherence scale can thus be defined as
\eq{
\SCS_n^2(\rho)\equiv&\frac{1}{{2}\mathcal{J}_n\mathcal{P}(\rho)}\sum_{i=1}^d\Tr([\rho,J_i][J_i,\rho])\\
=&\frac{1}{{2}\mathcal{J}_n}\sum_{i=1}^d\sum_{mm^\prime}(m-m^\prime)^2 P_\rho(m,m^\prime;i),
} where now we must use the eigenbases of the SU($n$) generators when we define the coherences' probability distributions as in Eq.~\eqref{eq:coherence prob dist} with appropriate basis states $\{|\mathcal{J}_n m\rangle_i\}$. We also now normalize by an appropriate quantity $\mathcal{J}_n$ that will make the classical/quantum threshold again occur at $\SCS_n=1$. In the case of SU(2), the normalization constant $J$ is the spin that is in one-to-one correspondence with the quadratic Casimir invariant $J(J+1)$ and is uniquely given by $J=N/2$ when the spin is made from a symmetric combination of $N$ 2-level particles. The spin $J$ also uniquely determines the irreducible representation of SU(2) under consideration. 

For SU($n$), the irreducible representations require multiple parameters to be specified and each have their own coherent states~\cite{Nemoto2000,MathurRaychowdhury2011,ZhangBatista2021}. We focus on the physically relevant case of symmetric combinations of $N$ $n$-level particles, each of which is equivalent to a spin-$(n-1)/2$ system, such as $N$ photons arrayed among $n$ modes; this is the case in SU(3)~\cite{GnutzmannKus1998,Roweetal1999,MathurSen2001,NemotoSanders2001,ChaturvediMukunda2002} when investigating three-dimensional polarization properties of light~\cite{Luis2005su3coherent,Luis2005polarization3D}. This representation is labeled by $(N,0,0,\cdots)$ and thus requires only a single parameter to be specified. Then, the quadratic Casimir invariant can be read off from $\sum_{i=1}^d J_i^2=C_n(N)\mathbb{I}$ to be $C_n(N)=N(N+n)(n-1)/2n$ and we subsequently find $\mathcal{J}_n=N(n-1)/2$.
Physically, the parameter $N$ sets the energy or the number of resources of the SU($n$) system in question.

Since the generators have good transformation properties under the group, we can show this coherence scale to be invariant under SU$(n)$ operations. We do this by extending the case of rotations to SU($n$) unitaries $U(\theta,\mathbf{n})=\exp(\iu\theta\mathbf{J}\cdot\mathbf{n})$ for $\mathbf{J}=(J_1,\cdots,J_d)^\top$ and $d$-dimensional unit vectors $\mathbf{n}$, where now the generators transform via
\eq{
U(\theta,\mathbf{n})J_iU(\theta,\mathbf{n})^\dagger=\sum_{j=1}^du_{ij}(\theta,\mathbf{n}) J_j
} using the elements $u_{ij}(\theta,\mathbf{n})$ of a $d\times d$ unitary matrix. In fact, since each generator is Hermitian, the right-hand side is equivalent to $\sum_{j=1}^du_{ij}(\theta,\mathbf{n})^* J_j$. The SU($n$) coherence scale simplifies to
\eq{
\SCS_n^2=\frac{C_n(N)}{\mathcal{J}_n}-\frac{1}{\mathcal{J}_n\mathcal{P}(\rho)}\sum_{i=1}^d\Tr(J_i\rho J_i\rho),
} from which we immediately see the invariance
\eq{
\sum_{i=1}^d\Tr(J_iU^\dagger\rho U  J_iU^\dagger\rho U)&=
\sum_{i,j,k=1}^d \Tr(u_{ij}^*J_j \rho u_{ik}J_k\rho )\\
&=
\sum_{i}^d \Tr(J_i \rho J_i\rho )
} so that $\SCS_n^2(U(-\theta,\mathbf{n})\rho U(\theta,\mathbf{n}))=\SCS_n^2(\rho)$.

The ability to write this coherence scale as minus the time rate of change of purity again follows from the depolarization channel whose $d$ Lindblad operators are all of the generators $J_i$, which is also the channel describing a continuous measurement of all of the $\{J_i\}$~\cite{Barchiellietal1982,CavesMilburn1987}. This again has the unique steady state of the maximally mixed state, is a unital evolution, and so on. We present in Appendix~\ref{app:suN invariant depol} a proof that this unique SU($n$)-invariant depolarization channel indeed takes the form
\eq{
\frac{\partial \rho}{\partial t}\propto \sum_{i=1}^d [J_i,[J_i,\rho]],
}
which itself could be the subject of future investigation. For pure states the {coherence} scale again becomes a sum of variances of all of the generators
. The SU($n$) coherent states of the $(N,0,0,\cdots)$ irreducible representation can be defined as the most classical, which happen to be the ones that maximize $\sum_i\langle J_i\rangle^2$ and are all related to each other by an SU($n$) unitary, each achieving $\SCS_n^2=1$. 
Convex combinations of such classical states can only have smaller coherence scales as can be seen by choosing an appropriate basis along the lines of Eqs.~\eqref{eq:off diag coherent state inequality}~and~\eqref{eq:convex coherent inequality}, making any state with an SU($n$) coherence scale greater than that of a coherent state manifestly nonclassical. The scale can thus be used to again bound the distance to the set of classical states from both above and below.

All of the connections above should hold, too. The connection to metrology via pure states is upheld: the SU($n$) coherence scale is proportional to the QFI for estimating the angle $\theta$ of an SU($n$) unitary when averaged over all axes $\mathbf{n}$ and for the trace of the QFI when estimating all $d$ parameters of the unitary with a parametrization-independent weighting~\cite{Goldbergetal2021intrinsic}. Monotonicity under noise for the purity is immediate from the same 2-quasi-relative entropy computation. The only remaining unproven aspects are those requiring $s$-ordered quasiprobability distributions for SU($n$), as they are beyond our scope. \sugg{Can one bound negativity in SU($n$) systems with the SU($n$) coherence scale?} Do the coefficients of the irreducible tensor operators that arise in the $s$-ordered quasiprobability distributions for SU($n$)~\cite{KlimovdeGuise2010,TilmaNemoto2012} obey useful asymptotic properties in $N$, as do the Clebsch-Gordan coefficients in Eq.~\eqref{eq:s ordered SU2} for SU(2)? Should they enjoy the same properties as SU(2), with the multipole moments decaying exponentially and the quasiprobability distributions evolving over time to ones with smaller $s$, then all of the properties proven for the QCS and $\SCS^2$ will be seen to be faithfully instantiated throughout SU($n$).

\section{Conclusions}
The spin coherence scale connects the amount of coherence present in a state to the noncommutative nature of angular momentum operators, to a state's usefulness for single- and multiparameter rotation sensing, to spin squeezing, to nonclassicality witnessing and distance measures, and to loss of purity with depolarization noise. All of these properties also hold for generic physical systems governed by SU($n$) \sugg{and, since the quadrature coherence scale naturally generalizes to multiple bosonic modes, we speculate that the spin coherence scale may be generalized to composite systems}.
We have thus shown the quantumness properties of ambiguous physical systems to be coherent.

\begin{acknowledgments}
\sugg{AZG and AH} acknowledge that the NRC headquarters is located on the traditional unceded territory of the Algonquin Anishinaabe and Mohawk people, as well as support from NRC's Quantum Sensors Challenge Program. \sugg{This work was supported by NSERC under Discovery Grant RGPIN-2020-05767, the QuEnSi quantum alliance (NSERC ALLRP 578468 - 22), and the John Templeton Foundation under grant ID 63209. Additional support came from the Fetzer Franklin Fund of the John E. Fetzer Memorial Trust. AMS is a fellow of CIFAR.} 
\end{acknowledgments}

\clearpage
\onecolumngrid

\appendix


\section{Positivity Threshold Values}
\label{app:pos thresholds}
Table~\ref{tab} lists the values where families of states lose their Wigner negativity and where our bounds can certify Wigner non-negativity.
\begin{table*}[h]
\caption{Numerical values of the bound-specified positivity threshold $p_{\mathrm{pos}}$ and the exact Wigner-positivity threshold $p_{\mathrm{crit}}$ for the state families and spin values shown in Fig. \ref{fig:scs depol}. Values are rounded to 5 decimal places; pure spin coherent states never achieve nonnegativity at $p = 1$.}
\begin{ruledtabular}
\begin{tabular}{lccc}
State & $J$ & $p_{\mathrm{pos}}$ & $p_{\mathrm{crit}}$ \\
Spin Coherent & 3  & 0.07674 & 0.98060 \\
Spin Coherent & 10 & 0.01951 & 1.00000 \\
Spin Coherent & 20 & 0.00891 & 1.00000 \\
NOON & 3  & 0.03837 & 0.18106 \\
NOON & 10 & 0.00585 & 0.08780 \\
NOON & 20 & 0.00183 & 0.05499 \\
Berry--Wiseman & 3  & 0.07417 & 0.73828 \\
Berry--Wiseman & 10 & 0.01440 & 0.51860 \\
Berry--Wiseman & 20 & 0.00495 & 0.35913 \\
Yurke & 3  & 0.03922 & 0.20604 \\
Yurke & 10 & 0.00585 & 0.15492 \\
Yurke & 20 & 0.00191 & 0.08762 \\
\end{tabular}
\end{ruledtabular}
\label{tab}
\end{table*}

\section{Axisymmetry of Bound-Saturating States}
\label{app:axisymmetry}

In this section, we show that all states saturating the fluctuation bound are axisymmetric up to rotation.
From the extremality conditions derived in the main text, the multipole coefficients of any saturating state take the form
\begin{equation}
\rho_{Kq}
=
\lambda \frac{Y_{Kq}^*(\Omega_*)}{K(K+1)},
\label{eq:extremal_form_appendix}
\end{equation}
for some location $\Omega_*$ on the sphere.

Let $R$ be a rotation that maps the north pole $\Omega_0$ to $\Omega_*$. Under rotations, spherical harmonics transform as
\begin{equation}
Y_{Kq}(\Omega_*)
=
\sum_{q'=-K}^{K} D^{(K)}_{q q'}(R)\, Y_{Kq'}(\Omega_0),
\end{equation}
where $D^{(K)}_{q q'}(R)$ are Wigner $D$-matrix elements.

At the north pole, only the $q'=0$ component is nonzero:
\begin{equation}
Y_{Kq'}(\Omega_0)
=
\delta_{q'0}\,\sqrt{\frac{2K+1}{4\pi}}.
\end{equation}
Thus,
\begin{equation}
Y_{Kq}(\Omega_*)
=
D^{(K)}_{q0}(R)\,\sqrt{\frac{2K+1}{4\pi}}.
\end{equation}

Substituting into Eq.~\eqref{eq:extremal_form_appendix}, we obtain
\begin{equation}
\rho_{Kq}
=
\lambda\, D^{(K)}_{q0}(R)\,
\frac{\sqrt{(2K+1)/(4\pi)}}{K(K+1)}.
\end{equation}

Now define the reference multipole coefficients
\begin{equation}
\rho^{(0)}_{Kq}
=
\delta_{q0}\,
\lambda
\frac{\sqrt{(2K+1)/(4\pi)}}{K(K+1)}.
\end{equation}
Under a rotation $R$, multipole coefficients transform as
\begin{equation}
\rho_{Kq}
=
\sum_{q'=-K}^{K} D^{(K)}_{q q'}(R)\,\rho^{(0)}_{Kq'}.
\end{equation}
Since $\rho^{(0)}_{Kq'}$ is nonzero only for $q'=0$, this reduces to
\begin{equation}
\rho_{Kq}
=
D^{(K)}_{q0}(R)\,\rho^{(0)}_{K0},
\end{equation}
which coincides with the extremal form above.

Therefore, any bound-saturating state can be obtained by rotating a state whose multipole coefficients are supported entirely on the $q=0$ sector. Such states are axisymmetric about a fixed axis.

\begin{equation}
\boxed{
\text{All bound-saturating states are axisymmetric up to rotation.}
}
\end{equation}

\section{Derivations for SU($n$)}
\label{app:suN details}
One physical manifestation of SU($n$) is an $n$-mode system with $N$ photons. This system will be sufficient for proving all of our properties of SU($n$) in the $(N,0,0,\cdots)$ irreducible representation, which can then include other physical systems such as symmetric states of $N$ spin-$(n-1)/2$
particles.

Consider the $n$ bosonic annihilation operators $a_i$. With these we can form the generators of the algebra:
\eq{
\{J_i\}=\left\{\left\{\frac{a^\dagger_i a_j+a_j^\dagger a_i}{2}\right\}_{1\leq i<j\leq n}
,
\left\{\frac{a^\dagger_i a_j-a_j^\dagger a_i}{2\iu}\right\}_{1\leq i<j\leq n}
,\left\{\frac{1}{\sqrt{2k(k-1)}}\left((1-k)a_k^\dagger a_k+\sum_{i=1}^{k-1}a_i^\dagger a_i\right)\right\}_{1<k\leq n}\right\}.
\label{eq:suN generators bosonic}
} The first two sets of terms generalize $J_1$ and $J_2$ and together form ladder operators like $a_i^\dagger a_j$ while the third set involves only photon-number operators like $a_i^\dagger a_i$ and generalizes $J_3$. These are found via the Jordan map from generalized Gell-Mann matrices to Fock space. Unitaries formed by exponentiating these generators preserve photon number, so we restrict to a fixed irreducible representation of SU($n$) with $N$ photons, spanned by the states
\eq{
|\mathbf{m}\rangle=\prod_{i=1}^n \frac{a_i^{\dagger m_i}}{\sqrt{m_i}}|\mathrm{vac}\rangle,\qquad \sum_{i=1}^n m_i=N.
}

Just like for SU(2), the quadratic Casimir invariant is uniquely specified by $N=2J$ once we restrict to a particular type of irreducible representation. Unlike SU(2), however, it is not simply equal to $\tfrac{N}{2}(\tfrac{N}{2}+1)$. Instead, a standard computation using $(a^\dagger_i a_j+a_j^\dagger a_i)^2-(a^\dagger_i a_j-a_j^\dagger a_i)^2=4 a_i^\dagger a_i a_j^\dagger a_j+2a_i^\dagger a_i+2a_j^\dagger a_j$ and $d=n^2-1$
yields
\eq{
\sum_{i=1}^d J_i^2=\frac{N(N+n)(n-1)}{2n}\mathbb{I}_N\equiv C_n(N) \mathbb{I}_N
\label{eq:Casimir suN}
} when acting on the $N$-photon subspace and, therefore, $\sum_{i=1}^d J_i^2=\sum_{N>0}C_n(N)\mathbb{I}_N$ when acting on a state with support on multiple photon-number subspaces. One verifies $C_2(2J)=J(J+1)$ and that $C_n$ is quadratic in $N$ such that an appropriate reference for determining an energetic resource is often $\sqrt{C_n(N)}$ in the large-$N$ limit.

The Casimir invariant is almost enough to specify the coherence scale:
\eq{
\SCS^2_n=\frac{C_n(N)}{\mathcal{J}_{n}}-\frac{1}{\mathcal{J}_{n}\mathcal{P}}\sum_{i=1}^d\Tr((\rho J_i)^2) ,
} 
where $\mathcal{J}_n$ is a normalization constant that sets the threshold between the classical and quantum regimes.

For pure states, the latter term is $-1/\mathcal{J}_n$ times
\eq{
\sum_{i=1}^d\langle J_i\rangle^2= \frac{n-1}{2n}\sum_{i=1}^n\langle a_i^\dagger a_i\rangle^2+\frac{1}{2}\sum_{i\neq j}|\langle a_i^\dagger a_j\rangle|^2-\frac{\langle a_i^\dagger a_i\rangle \langle a_j^\dagger a_j\rangle}{n}\leq \frac{n-1}{2n}\left(\sum_{i=1}^n\langle a_i^\dagger a_i\rangle\right)^2=\frac{n-1}{2n}N^2.
} The inequality follows from repeated application of the Cauchy-Schwarz inequality $|\langle a_i^\dagger a_j\rangle|^2\leq \langle a_i^\dagger a_i\rangle \langle a_j^\dagger a_j\rangle$, used in Ref.~\cite{Luis2005su3coherent} to identify the SU(3) coherent states. It is then immediate that the pure states with the smallest coherence scale are the SU($n$) coherent states that saturate all of the inequalities by obeying 
\eq{
a_i|\theta,\mathbf{n}^{(n;N)}\rangle\propto a_j|\theta,\mathbf{n}^{(n;N)}\rangle\qquad \forall i,j.
}

We preemptively labeled the coherent states by $\theta$, $\mathbf{n}$, $n$, and $N$. The latter two are necessary for specifying the group and the irrep to which the coherent states belong, but why the other two symbols, other than generalizing the angular coordinates $\Omega$? A glance shows that one particular state satisfies all of the Cauchy-Schwarz inequalities: $|N,0,\cdots,0\rangle$, for which $n-1$ of the annihilation operators simply annihilate the state. What then follows is that any SU($n$) transformation of this state is also an SU($n$) coherent state, which can be seen either by all coherent states being related by group displacements~\cite{Perelomov1986,Nemoto2000} or by the proof in the main text that $\sum_i\langle J_i\rangle^2$ is unchanged by SU($n$) operations, such that the states maximizing this quantity are all of the form \eq{
U(\theta,\mathbf{n})|N,\underbrace{0,\cdots,0}_{n-1\,\mathrm{times}}\rangle\equiv |\theta,\mathbf{n}^{(n;N)}\rangle.
}

For the coherent states, the coherence scale is exactly $\SCS_n^2=N(n-1)/2\mathcal{J}_{{n}}$, which can be computed by adding up $\sum_i\langle N,0,\cdots, 0|J_i|N,0,\cdots,0\rangle^2=N^2(n-1)/2n$. Since the threshold for nonclassicality is exactly met for coherent states, one could set $\mathcal{J}_{{n}}=N(n-1)/2$ to continue the trend of $\SCS_n^2>1$ heralded quantumness. To prove that indeed all convex combinations of coherent states yield $\SCS_n(\rho_{\mathrm{cl}})\leq 1$, we invoke a similar argument to Eq.~\eqref{eq:off diag coherent state inequality} to show that \eq{
\sum_{i=1}^d |\langle \theta_1,\mathbf{n}_1^{(n;N)}|J_i|\theta_2,\mathbf{n}_2^{(n;N)}\rangle|^2=\sum_{i=1}^d |\langle \theta_1,\mathbf{n}_1^{(n;N)}|U(\theta_2,\mathbf{n}_2)^\dagger J_i|N,0\cdots ,0\rangle|^2\geq \frac{n-1}{2n}N^2|\langle \theta_2,\mathbf{n}_1^{(n;N)}|\theta_2,\mathbf{n}_2^{(n;N)}\rangle|^2
,
} where equality holds if one restricts the sum to the third set of operators in Eq.~\eqref{eq:suN generators bosonic}. Then the exact same logic as in Eq.~\eqref{eq:convex coherent inequality} but with $J(J+1)$ and $J$ upgraded to $C_n(N)$ and $\mathcal{J}_n$, respectively, in the definition of the coherence scale proves
\eq{
\SCS_n\left(\sum_k q_k | \theta_k,\mathbf{n}_k^{(n;N)}\rangle\langle \theta_k,\mathbf{n}_k^{(n;N)}|\right)\leq 1\qquad \forall q_k>0.
}

\section{SU($n$)-invariant depolarization channel}
\label{app:suN invariant depol}
Depolarization channels exist in all dimensions and eventually take any input state to the maximally mixed state. They physically arise from the average of many seemingly random operations acting on a state, which often come from ignorance and of small unknown effects. We here seek an additional property of SU($n$) invariance: for consistency, rotating a state prior to a depolarization channel should be equivalent to rotating the state after the depolarization channel, given that the maximally mixed state is unchanged by unitary operations. We follow Ref.~\cite{RivasLuis2013}'s pioneering study of SU(2)-invariant depolarization channels to show how very similar structures lead to SU($n$)-invariant depolarization channels in arbitrary dimensions.

A state subject to random SU($n$) operations (as usual, with $d=n^2-1$ generators) takes the form
\eq{
\rho(t)=\mathcal{E}_t(\rho(0))=\int \theta^{d-1}d\theta d\mathbf{n} p(\theta,\mathbf{n};t)U(\theta,\mathbf{n})\rho(0)U(\theta,\mathbf{n})^\dagger
} for a normalized, time-dependent probability distribution $p(\theta,\mathbf{n};t)$ subject to the initial condition $p(\theta,\mathbf{n};0)\propto \delta(\theta)$. The integration measure $\theta^{d-1}d\theta d\mathbf{n}$ is assumed to be normalized and takes that form because we have separated the length degree of freedom $\theta$ from the angular degrees of freedom $\mathbf{n}$. For this to be SU($n$)-invariant, we require \eq{
\mathcal{E}_t(V\rho(0)V^\dagger)=V\mathcal{E}_t(\rho(0))V^\dagger
} for any SU($n$) operation $V$, which implies that $p(\theta,v\mathbf{n};t)=p(\theta,\mathbf{n};t)$ for all SU($n$) matrices $v$ and thus that $p(\theta,\mathbf{n};t)=p(\theta;t)$. I.e., the probability of a particular operation must always be isotropic. 

Then, to be continuous at small $t$, we expect $p(\theta,t)$ to only have support over small values of $\theta$ when $t\ll 1$. This leads to an expansion of the density matrix using second-order approximations in $\theta$, $U(\theta,\mathbf{n}){=\exp(\iu\theta\mathbf{J}\cdot\mathbf{n}) }\approx \mathbb{I}{+}\iu \theta\mathbf{J}\cdot\mathbf{n}-\frac{\theta^2}{2}(\mathbf{J}\cdot\mathbf{n})^2$, and leads to many terms with vanishing angular integrals $\int d\mathbf{n} n_i =0$ and $\int d\mathbf{n} n_i n_j\propto \delta_{ij}$. Defining $t p^\prime(\theta)=p(\theta;t)-p(\theta;0)$ and $\int \theta^{d-1} d\theta d\mathbf{n}  p^\prime (\theta) n_i^2=\nu>0$, we find
\eq{
\rho(t)\approx  \rho(0)+t\nu \sum_{i=1}^d \left(J_i \rho(0)J_i-\frac{J_i^2\rho(0)+\rho(0) J_i^2}{2}\right),
}
which for small $t$ implies a master equation with Lindblad operators $J_i$ multiplied by a constant $\sqrt{\nu}$  that simply rescales the time coordinate.
Since $\rho(0)$ can be any input state, this defines the master equation for all time:
\eq{
\frac{\partial \rho}{\partial t}=-\frac{\nu}{2}\sum_{i=1}^d [J_i,[J_i,\rho]].
}

The SU($n$)-invariant depolarization channel may be compared to the standard depolarization channel $\sum_i K_i \rho K_i^\dagger=(1-p)\rho+p\mathbb{I}/D$ for states in dimension $D$. The latter has Kraus operators $K_i$ that may be expressed in many equivalent forms, one of which is $\{K_i\}=\{\sqrt{1-p}\mathbb{I},\{\sqrt{p/D}||m\rangle\langle m^\prime||\}\}$ for some orthonormal basis $\{||m\rangle\}$ where $m$ and $m^\prime$ range from $1$ to $D$. In the fundamental or defining representation of SU($n$), the two depolarization channels are equivalent, as follows. The fundamental representation is $(1,0,0\cdots)$ and represents an $n$-mode system with a single photon; there, importantly, $D=n$. The Kraus operator channel has $\tfrac{\partial \rho}{\partial p}=-\rho+\mathbb{I}/n=-\rho+\sum_{mm^\prime}||m\rangle\langle m^\prime||\rho||m^\prime\rangle\langle m||/n$, while the SU($n$)-invariant channel has the time evolution
\eq{
\frac{\partial \rho(t)}{\partial t}=-\rho(t)C_n(1)+\sum_{i=1}^d J_i\rho(t)J_i,
} which looks like it has Kraus operators $\sqrt{C_n(1)}\mathbb{I}$ and $\{J_i\}$ and some possibly nonlinear scaling between $t$ and $p$. Two sets of Kraus operators are equivalent if $K_i^\prime=\sum_j u_{ij}K_j$ for any unitary matrix with elements $u_{ij}$. We simply show that, in the space of $n$-mode single-photon states labeled using $||i\rangle$ to denote the single photon being in mode $i$, 
\eq{
\{J_i\}=\left\{\left\{\frac{||i\rangle\langle j||+||j\rangle\langle i||}{2}\right\}_{1\leq i<j\leq n}
,
\left\{\frac{||i\rangle\langle j||-||j\rangle\langle i||}{2\iu}\right\}_{1\leq i<j\leq n}
,\left\{\frac{1}{\sqrt{2k(k-1)}}\left((1-k)||k\rangle\langle k||+\sum_{i=1}^{k-1}||i\rangle\langle i||\right)\right\}_{1<k\leq n}\right\}
.} This is the set of generalized Gell-Mann matrices, which are all orthogonal according to the Hilbert-Schmidt inner product and together with the identity matrix span the vector space of $n\times n$ matrices. Therefore, after accounting for the normalization $\Tr[J_i J_k]=\delta_{ij}/2$, 
\eq{
\sum_{mm^\prime} ||m\rangle\langle m^\prime||\rho||m^\prime\rangle\langle m||=(\mathbb{I}/\sqrt{n})\rho(\mathbb{I}/\sqrt{n})+\sum_{i=1}^d (\sqrt{2}J_i)\rho (\sqrt{2}J_i)\quad \Rightarrow\quad \sum_i J_i\rho J_i=\frac{\mathbb{I}}{2}-\frac{\rho}{2n}.
} Then, finding from Eq.~\eqref{eq:Casimir suN} that $C_n(1)+1/2n=n/2$, we see that the SU($n$)-invariant depolarization channel is the standard depolarization channel when acting on the fundamental representation of SU($n$) subject to the rescaled time $\partial p/\partial t=n/2$; it generalizes this behaviour to other irreducible representations by not adding more Kraus operators and instead maintaining the structure of using SU($n$) generators as Kraus operators.

\end{document}